\documentclass{book}
\usepackage{makeidx,epsfig}
\usepackage{amsthm,amsmath,amssymb}
\usepackage{setspace,graphicx,xcolor}
\usepackage{Generic}
\usepackage[sort,longnamesfirst]{natbib}

\usepackage{natbib}

\usepackage{url}
\newcommand{\PFCSExpect}[1]{\mathbb{E}\left[{#1}\right]}
\newcommand{\PFCSExpects}[2]{\mathbb{E}_{{#1}}\left[{#2}\right]}
\newcommand{\PFCSmd}{\mathsf{d}}
\newcommand{\PFCSwtil}{\widetilde{w}}
\newcommand{\PFCSxtil}{\widetilde{x}}
\newcommand{\PFCSpihat}{\widehat{\pi}}

\newcommand{\PFCSphat}{\widehat{p}}

\numberwithin{equation}{section}
\theoremstyle{plain}

\begin{document}
\begin{doublespace}
%


\chapter[State Space Models]{MCMC for State Space Models}


\begin{center}
\begin{large}
{\em Paul Fearnhead and Chris Sherlock}
\end{large}
\end{center}

\vspace{.5cm}
\begin{center}
    \textbf{Abstract}
\end{center}
\vspace{0.1cm}

A state-space model is a time-series model that has an unobserved latent process from which we take noisy measurements over time. The observations are conditionally independent given the latent process and the latent process itself is Markovian. These properties lead to simplifications for the conditional distribution of the latent process given the parameters and the observations. This chapter looks at how we can leverage the properties of state-space models to construct efficient MCMC samplers. We consider a range of Gibbs-sampler schemes, including those which use the forward-backward algorithm to simulate from the full conditional of the latent process given the parameters. For models where the forward-backward algorithm is not applicable we look at particle MCMC algorithms that, given the parameters, use particle filters to approximately simulate from the latent process or estimate the likelihood of the observations. Throughout, we provide intuition and informally discuss theory about the properties of the model that impact the efficiency of the different algorithms and how approaches such as reparameterization can improve mixing.


\section{Introduction: State-space models} 
\label{s:intro}
In this chapter we look at MCMC methods for a class of time-series models, called state-space models. The idea of state-space models is that there is an unobserved state of interest which evolves through time, and that partial observations of the state are made at successive time-points. We will denote the state by $X$ and observations by $Y$, and assume that our state space model has the following structure:
\begin{equation}
X_t |\{x_{1:t-1},y_{1:t-1}\} \sim p(x_t|x_{t-1},\theta), \label{eq:SSM1}
\end{equation}
\begin{equation}
Y_t|\{x_{1:t},y_{1:t-1}\} \sim p(y_t|x_t,\theta). \label{eq:SSM2}
\end{equation}
Here, and throughout, we use the notation $x_{1:t}=(x_1,\ldots,x_t)$, and write $p(\cdot|\cdot)$ for a generic conditional probability density or mass function (with the arguments making it clear which conditional distribution it relates to). 
To fully define the distribution of the hidden state we further specify an initial distribution $p(x_1|\theta)$.
We have made explicit the dependence of the model on an unknown parameter $\theta$, which may be multi-dimensional.  The assumptions in this model are that, conditional on the parameter $\theta$, the state model is Markov, and that we have a conditional independence property for the observations: observation $Y_t$ only depends on the state at that time, $X_t$. Since the state is not observed, these models are often called \emph{hidden Markov models} (HMMs).

For concreteness we give two examples of state-space models and briefly list some others:

\noindent{\bf Example 1: Stochastic Volatility}

The following simple stochastic volatility (SV) model has been used for modelling the time-varying variance of log-returns on assets. For fuller details see \citet{Hull/White:1987} and \citet{Shephard:1996}. The state-space model is
\[
X_t|\{x_{1:t-1},y_{1:t-1}\} \sim \mbox{N}(\phi x_{t-1},\sigma^2),
\]
where $|\phi|<1$, and with initial distribution $X_1\sim\mbox{N}(0,\sigma^2/(1-\phi^2))$, and
\[
Y_t|\{x_{1:t},y_{1:t-1}\} \sim \mbox{N}(0,\beta^2\exp\{x_t\}).
\]
The parameters of the model are $\theta=(\beta,\phi,\sigma)$. The idea of the model is that the variance of the observations depends on the unobserved state, and the unobserved state is modelled by an AR(1) process.

\noindent{\bf Example 2: Discrete Hidden Markov Model}

A general class of models occurs when the underlying state is a discrete-valued Markov model, with a finite state-space. Thus we can assume without loss of generality that $X_t\in\{1,2,\ldots,K\}$ and that the model for the dynamics of the state (\ref{eq:SSM1}) is defined by a $K\times K$ transition matrix $P$. Hence, for all $i,j \in \{1,\ldots,K\}$:
\begin{equation} \label{eq:SSM3}
\Pr(X_t=j|X_{t-1}=i,x_{1:t-2},y_{1:t-1})=P_{ij}.
\end{equation}
Often, it is assumed that the distribution for $X_1$ is given by the stationary distribution of this Markov chain.
The observation equation (\ref{eq:SSM2}) will depend on the application, but there will be $K$ observation regimes (depending on the value of the state). Thus we can write
\begin{equation} \label{eq:SSM4}
Y_t|\{x_t=k,x_{1:t-1},y_{1:t-1}\} \sim f_k(y_t|\theta).
\end{equation}
The parameters of this model will be the parameters of (\ref{eq:SSM4}) and the parameters of the transition matrix $P$.

Examples of such models include models of Ion-channels \citep{Ball:1992,Hodgson:1999}, DNA sequences \citep{Boys/Henderson/Wilkinson:2000}, speech \citep{Juang/Rabiner:1991} and discrete-time models for epidemics \citep[e.g.][]{LekFin2006}.

\noindent{\bf Other examples}

State-space models are common in diverse applications such as target tracking \cite[]{GSS1993}, ecology \cite[]{king2014statistical}, epidemics \cite[]{touloupou2020scalable} and econometrics \cite[]{Kitagawa:1987} amongst many others.  They also include more general classes of statistical models, such as those used for changepoint detection or data segementation \citep[e.g.][]{Didelot:2007,Fearnhead:2008}. The class also includes models where the $X$ process remains Markovian but evolves in continuous time; for example, a Markov jump process arising from  a reaction network \citep[e.g.][]{wilkinson2018stochastic}  or a diffusion process \cite[]{golightly2008bayesian}.  



\section{Bayesian analysis and MCMC framework}

Our aim is to perform Bayesian inference for a state-space model given data $y_{1:n}$. We assume a prior for the parameters, $p(\theta)$, has been specified, and we wish to obtain the posterior of the parameters $p(\theta|y_{1:n})$, or in some cases we may be interested in the joint distribution of the state and the parameters $p(\theta,x_{1:n}|y_{1:n})$.

How can we design an MCMC algorithm to sample from either of these posterior distributions? In both cases, this can be achieved using data augmentation \citep[e.g.][]{Hobert:2008}. That is we design a Markov chain whose state is $(\theta,X_{1:n})$, and whose stationary distribution is $p(\theta,x_{1:n}|y_{1:n})$; samples from the marginal posterior $p(\theta|y_{1:n})$ can be obtained from 
samples from $p(\theta,x_{1:n}|y_{1:n})$ by discarding the $x_{1:n}$ components.
The reason for designing an MCMC algorithm on this state-space is that, for state-space models of the form (\ref{eq:SSM1}--
\ref{eq:SSM2}), we can write down the stationary distribution of the MCMC algorithm up to proportionality:
\begin{equation} \label{eq:SSM5}
p(\theta,x_{1:n}|y_{1:n}) \propto p(\theta)p(x_1|\theta)\left(\prod_{t=2}^n p(x_t|x_{t-1},\theta) \right)
\left(\prod_{t=1}^n p(y_t|x_t,\theta) \right). 
\end{equation}

In most applications it is straightforward to implement an MCMC algorithm with (\ref{eq:SSM5}) as its stationary distribution. A common approach is to design moves that update $\theta$ conditional on the current values of $X_{1:n}$ and then update $X_{1:n}$ conditional on $\theta$. We will describe various approaches within this framework. We first focus on the problem of updating the state, considering $\theta$ fixed and known; we, then, examine moves to update the parameters. The remainder of this chapter then describes particle MCMC methods, which use particle filters to help create better-mixing MCMC moves.


\section{Updating the state}

The simplest approach to update the state $X_{1:n}$ is to update its components one at a time. Such a move is called a {\em single-site update}. While easy to implement, it can lead to slow mixing if there is strong temporal dependence in the
state process. In these cases it is better to update blocks of state components, $X_{t:s}$.
Sometimes, it is possible to update the whole process $X_{1:n}$ directly from its full-conditional distribution $p(x_{1:n}|y_{1:n},\theta)$, which can be  particularly effective.

We will give examples of single-site moves, and investigate when they do and do not work well, before looking at designing efficient block updates. For convenience we drop the conditioning on $\theta$ in the notation that we use within this section.

\subsection{Single-site updates of the state} \label{SSUP}
The idea of single-site updates is to design MCMC moves that update a single value of the state, $x_t$, conditional on all other values of the state process (and on $\theta$). Repeated application of this move for $t=1,\ldots,n$ will enable the whole state process to be updated.

We let $x_{-t}=(x_1,\ldots,x_{t-1},x_{t+1},\ldots,x_n)$ denote the whole state process excluding $x_t$, so a single-site update will update $x_t$ for fixed $x_{-t},\theta$. The target distribution of such a move is the full-conditional distribution $p(x_t|x_{-t},\theta,y_{1:t})$; which as mentioned above we will write as $p(x_t|x_{-t},y_{1:t})$ -- dropping the conditioning on $\theta$ in the notation that we use, as we are considering moves for fixed $\theta$. Due to the Markov structure of our model this simplifies to $p(x_t|x_{t-1},x_{t+1},y_{t})$  for $t=2,\ldots,n-1$, $p(x_1|x_2,y_1)$ for $t=1$ and $p(x_n|x_{n-1},y_n)$ for $t=n$.  Sometimes we can simulate directly from these full conditional distributions, and such ({\em{Gibbs}}) moves will always be accepted. Where this is not possible, then if $x_t$ is low-dimensional we can often implement an efficient {\em Independence Sampler} (see below).

We now detail single-site updates for Example 2 (Gibbs move) and Example 1 (Independence Sampler move), and in both cases we investigate the mixing properties for  updating $X_{1:n}$. 

\noindent{\bf Example 2: Single-site Gibbs move.} 

For the HMM model of Example 2, with state transition matrix, $P$, we have for $t=2,\ldots,n-1$ that
\begin{eqnarray*}
 \Pr(X_t=k|X_{t-1}=i,X_{t+1}=j,y_{t})&\propto& \Pr(X_t=k|X_{t-1}=i)\Pr(X_{t+1}=j|X_t=k)p(y_t|X_t=k)\\
&=& P_{ik}P_{kj}f_k(y_t),
\end{eqnarray*}
for $k=1,\ldots,K$. Now as $X_t$ has a finite state-space, we can calculate the normalising constant of this conditional distribution, leading to
\[
 \Pr(X_t=k|X_{t-1}=i,X_{t+1}=j,y_{t})=\frac{P_{ik}P_{kj}f_k(y_t)}{\sum_{l=1}^K P_{il}P_{lj}f_l(y_t)}.
\]
Similarly we obtain $\Pr(X_1=k|X_2=j,y_1)\propto \Pr(X_1=k)P_{kj}f_k(y_1)$ and $\Pr(X_n=k|X_{n-1}=i,y_n)\propto P_{ik}f_k(y_n)$. In both cases the normalising constants of these conditional distributions can be obtained.

Thus, for this model we can simulate from the full-conditionals directly, which is the optimal proposal for $x_t$ for fixed $x_{-t}$. Note that the computational cost of simulation is $O(K)$, due to calculation of the normalising constants. For large $K$ it may be more computationally efficient to use other proposals (such as an independence proposal) whose computational cost does not depend on $K$.

We examine the efficiency of this MCMC move at updating the state $X_{1:n}$ by focusing on a  HMM model for DNA sequences \citep[see e.g.][]{Boys/Henderson/Wilkinson:2000}. The data consists of a sequence of DNA, so $y_t\in\{\mbox{A,C,G,T}\}$ for all $t$. For simplicity we consider a two-state HMM, with the likelihood function for $k=1,2$ being
\[
 \Pr(Y_t=y|X_t=k)=\pi_{y}^{(k)}, \mbox{ for $y\in\{\mbox{A,C,G,T}\}$}.
\]
We denote the parameter associated with $X_t=k$ as $\pi^{(k)}=(\pi^{(k)}_{\mbox{\small{A}}},\pi^{(k)}_{\mbox{\small{C}}},\pi^{(k)}_{\mbox{\small{G}}},\pi^{(k)}_{\mbox{\small{T}}})$

We will consider the effect that both the dependence in the state dynamics, and the information in the observations have on the mixing rate of the MCMC move. To do this we will assume that state transition matrix satisfies $P_{12}=P_{21}=\alpha$, and
\[
\pi^{(1)}=\frac{1}{4}(1,1,1,1)+\beta(1,1,-1,-1) ~~~~ \pi^{(2)}=\frac{1}{4}(1,1,1,1)-\beta(1,1,-1,-1),
\]
for $0<\alpha<1$ and $0<\beta<1/4$. Small values of $\alpha$ correspond to large dependence in the state dynamics, and small values of $\beta$ correspond to less informative observations.

To measure the mixing properties of the single-site MCMC update we (i) simulated data for a given value of $(\alpha,\beta)$; (ii) ran an MCMC algorithm with single-site updates; and (iii) calculated an autocorrelation function for the MCMC output after discarding a suitable burn-in. For simplicity, we summarised the output based on the autocorrelation at lag-1 (all MCMC runs suggested autocorrelations that decayed approximately exponentially). We calculated the autocorrelation for the Hamming distance of the set of hidden states from the corresponding true states: the number of differences between the true value of the hidden state and the inferred value of the state.

\begin{figure}
\begin{center}
\rotatebox{0}{\includegraphics[height=4in,width=5in]{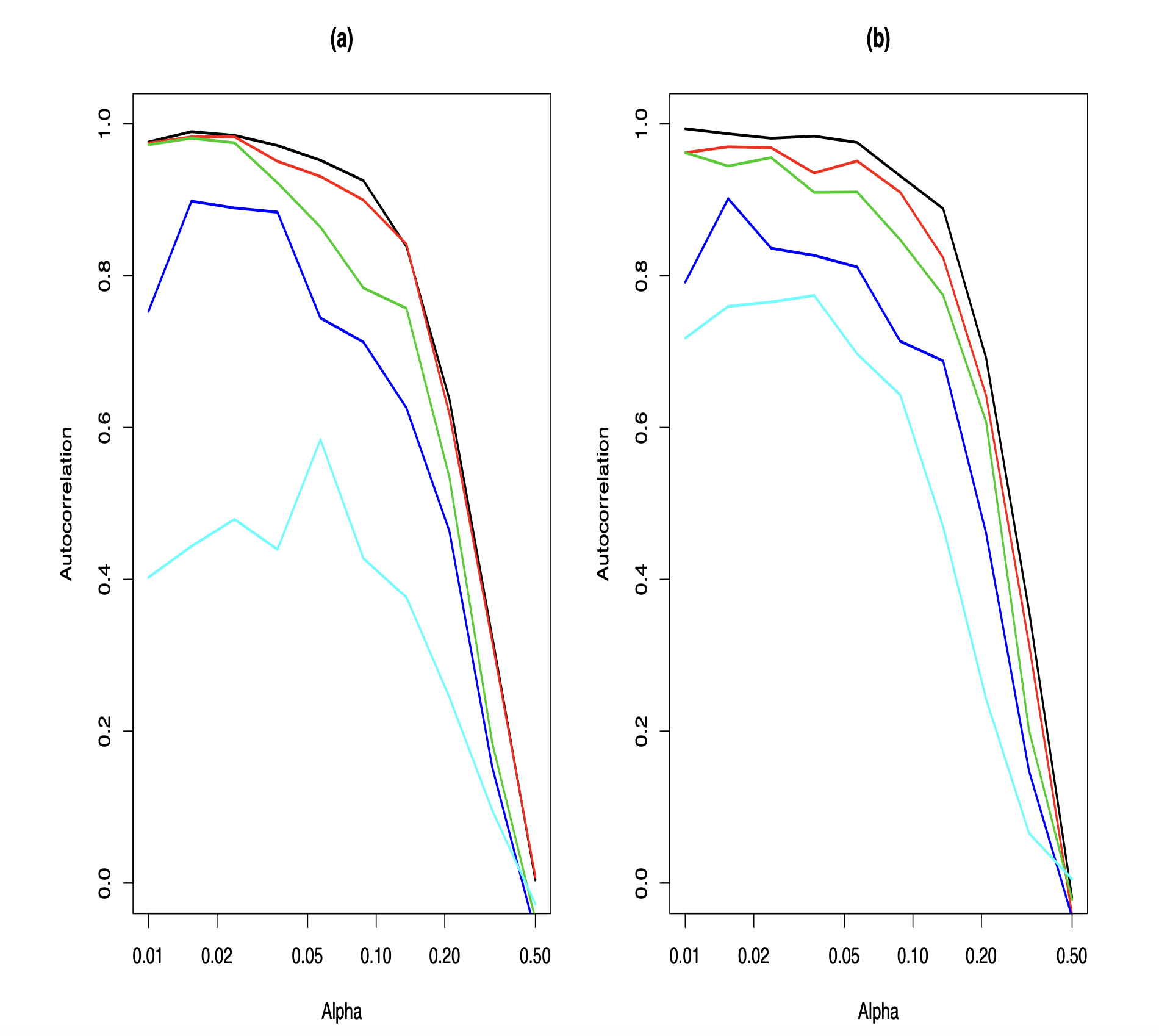} }
  \caption{\label{fig:HMMacf}Lag-1 autocorrelation of the Hamming distance between the hidden states and the truth for differing $\alpha$  for a 2-state HMM model: (a) $n=200$; (b) $n=500$.
In each plot, different lines refer to different values of $\beta$; from top to bottom: $\beta=0.02$ (black); $\beta=0.065$ (red);
$\beta=0.11$ (green); $\beta=0.155$ (dark blue); and $\beta=0.2$ (light blue). }
\end{center}
\end{figure}

Results are shown in Figure \ref{fig:HMMacf}, where we see that the value of $\alpha$ is the main determinant of the mixing of the MCMC algorithm. Small values of $\alpha$, which correspond to large dependence, result in poor mixing. Similarly, as $\beta$ decreases, which relates to less informative observations, the mixing gets worse -- though the dependence on $\beta$ is less than on $\alpha$. Qualitatively similar results are observed for the two values of $n$, but for smaller $n$ we see that the value of $\beta$ has more impact on the mixing properties.

\noindent{\bf Example 1: Single-site Independence Sampler}

Now consider the stochastic volatility (SV) model of Example 1. We describe an independence sampler that was derived by \citet{Shephard/Pitt:1997}. 

With this model, for $t=2,\ldots,n-1$ we obtain
\begin{eqnarray}
\lefteqn{ p(x_t|x_{t-1},x_{t+1},y_t) \propto p(x_t|x_{t-1})p(x_{t+1}|x_t)p(y_t|x_t)}  \nonumber \\
&\propto& \exp\left\{-\frac{1}{2\sigma^2}( (x_t-\phi x_{t-1})^2+(x_{t+1}-\phi x_t)^2 ) 
-\frac{x_t}{2} 
-\frac{\exp\{-x_t\}y_t^2}{2\beta^2} \right\}, \label{eq:SSM6}
\end{eqnarray}
where we have removed any constants of proportionality that do not depend on $x_t$; the first term in the exponent of the final expression corresponds to the two state transition densities, and the final two terms come from the likelihood.

Simulating directly from this conditional distribution is not possible, so we resort to approximation. Our approximation is based on a Taylor expansion of $\log p(x_t|x_{t-1},x_{t+1},y_t)$ about an estimate of $x_t$, which we call $\hat{x}_t$. Now if we define
$\mu_t=\phi(x_{t-1}+x_{t+1})/(1+\phi^2)$ and $\tau^2=\sigma^2/(1+\phi^2)$, then the first term in the exponent of  (\ref{eq:SSM6}) can be re-written as $-(x_t-\mu_t)^2/(2\tau^2)$. Thus without any observation, our conditional distribution of $x_t$ would have a mean $\mu_t$, and this appears a sensible value about which to take a Taylor expansion. Doing this we obtain
\[
 \log p(x_t|x_{t-1},x_{t+1},y_t) \approx -\frac{(x_t-\mu_t)^2}{2\tau^2}-\frac{x_t}{2}-\frac{y_t^2}{2\beta^2}\exp\{-\mu_t\}
\left(1-(x_t-\mu_t)+\frac{1}{2}(x_t-\mu_t)^2 \right).
\]
As this approximation to the log-density is quadratic, it gives us a Normal approximation to the conditional distribution, which
 we denote by $q(x_t|x_{t-1},x_{t+1},y_t)$. \citep[For full details of the mean and variance of the approximation, see][]{Shephard/Pitt:1997}. Thus we can implement an MCMC move of $X_t$ via an independence sampler with proposal $q(x_t|x_{t-1},x_{t+1},y_t)$.

Similar normal approximations can be obtained for $p(x_1|x_2,y_1)$ and $p(x_n|x_{n-1},y_n)$, the only difference is in the values of $\mu_t$ and $\tau$. Better estimates of $\hat{x}_t$ can be found, e.g. by numerically finding the mode of $p(x_t|x_{t-1},x_{t+1},y_t)$ \citep{Smith/Santos:2006}, but for single-site updates any increase in acceptance rate is unlikely to be worth the extra computation.

We investigate the efficiency of single-site updates for the SV model via simulation. We fix $\beta=1$ and consider how mixing of the MCMC algorithm depends on the time dependence of the state process, $\phi$, and marginal variance of the state process, $\tau^2=\sigma^2/(1-\phi^2)$. As above, we evaluate mixing by looking at the lag-1 autocorrelation of the mean squared error in the estimate of the state process.
\begin{figure}
\begin{center}
 \includegraphics[height=4in,width=4in,angle=0]{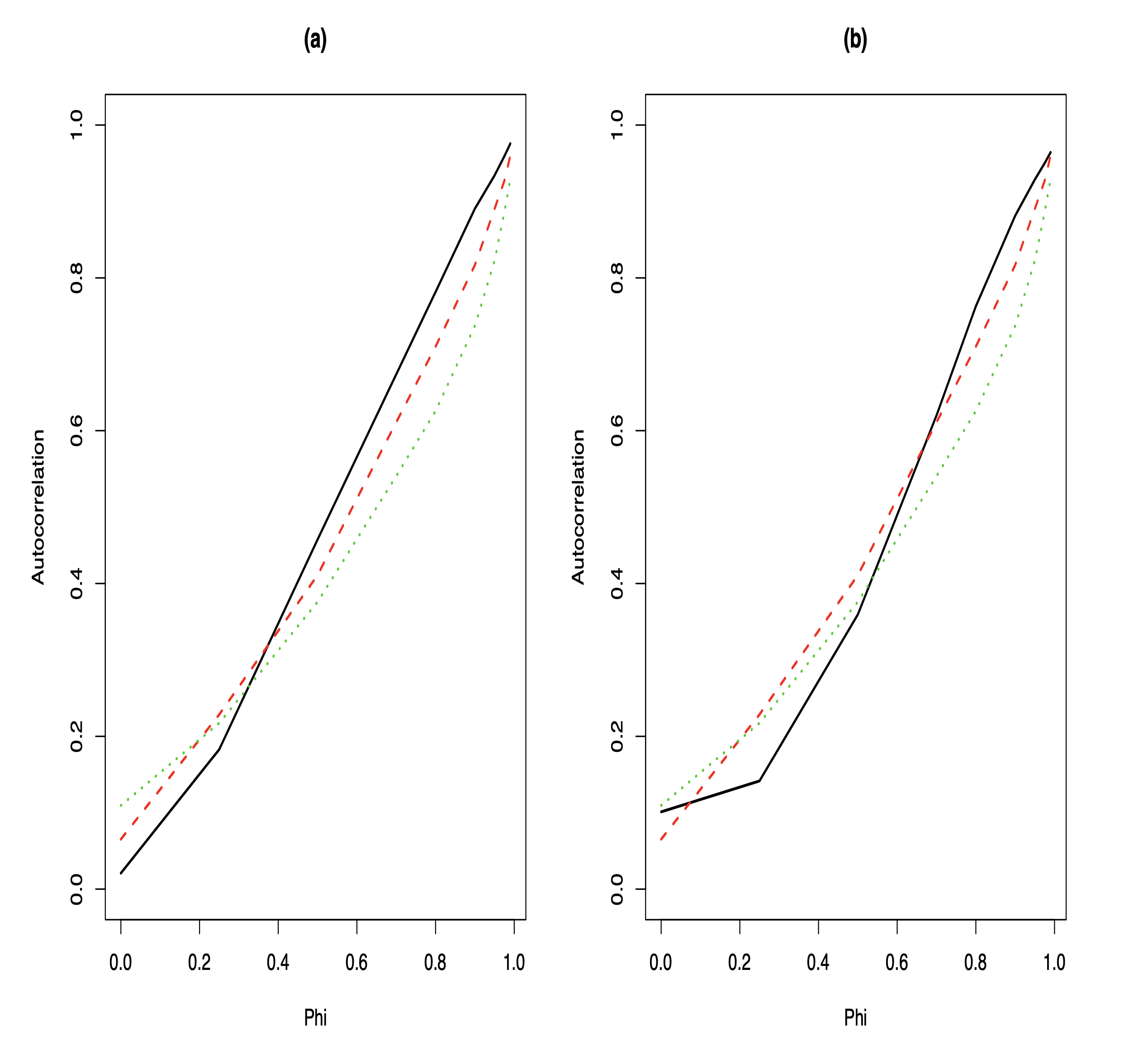}
 \caption{\label{fig:SVacf}Lag-1 autocorrelation of the mean squared error in the state process for differing $\phi$  for the SV model: (a) $n=200$; (b) $n=500$.
In each plot, different lines refer to different values of $\tau^2$: $\tau^2=0.5$ (black, full lines); $\tau^2=1$ (red, dashed lines); $\tau^2=2.0$ (green, dotted lines). }
\end{center}
\end{figure}
Results are shown in Figure \ref{fig:SVacf}, where we see that $\phi$ has a sizeable effect on mixing -- with $\phi\approx1$, which corresponds to strong correlation in the state process, resulting in poor mixing. By comparison both $n$ and $\tau^2$ have little effect. For all MCMC runs the acceptance rate of the MCMC move was greater than $99\%$.

\subsection{Block updates for the state}

While the single-site updates of Section \ref{SSUP} are easy to implement, we have seen that the resulting MCMC algorithms can mix slowly if there is strong dependence in the state-process. This leads to the idea of \emph{block updates}, which update the state at more than one time-point in a single move. Ideally we would update the whole state process in one move, and  in some cases it turns out that this is possible to do from the full-conditional, so that moves are always accepted. These cases include the linear-Gaussian model, where we can use the Kalman Filter \citep[see e.g.][]{Harvey:1989,Carter/Kohn:1994} and the HMM model of Example 2. We give details of the methods used for the latter case below. 

In situations where it is not possible to update the whole state process from its full conditional, one possibility is to use an independence proposal to update jointly a block of state values. We will describe such an approach for the SV model of Example 1; and then discuss alternative approaches for block updates for models where it is not possible to draw from the full conditional distribution of the state.

\noindent{\bf Example 2: Updating state from its full conditional}

The forward-backward algorithm is a method for sampling from the full conditional of the state-process for discrete HMMs. See \citet{Rabiner:1986} for a review of this method, and \citet{Scott:2002a} for further examples of its use within Bayesian inference. Here we describe its implementation for the model of Example 2.

The algorithm is based upon a forward recursion which calculates the filtering densities $\Pr(X_t|y_{1:t})$ for $t=1,\ldots,n$; followed by a backward simulation step that simulates from $\Pr(X_n|y_{1:n})$ and then $\Pr(X_t|y_{1:n},x_{t+1})$ for $t=n-1,\ldots,1$. The forward recursion is initialised with 
\[
 \Pr(X_1=k|y_1)\propto \Pr(X_1=k)f_k(y_1), \mbox{ for $k=1,\ldots,K$},
\]
where the normalising constant is $p(y_1)=\sum_{l=1}^K \Pr(X_1=l)f_l(y_1)$. Then for $t=2,\ldots,n$ we have
\[
 \Pr(X_t=k|y_{1:t})\propto f_k(y_t) \sum_{l=1}^K \Pr(X_{t-1}=l|y_{1:t-1})P_{lk}, \mbox{ for $k=1,\ldots,K$},
\]
where the normalising constant is $p(y_t|y_{1:t-1})$. A by-product of the forward recursions is that we obtain the likelihood as a product of these normalising constants, as $p(y_{1:n})=p(y_1)\prod_{t=2}^n p(y_t|y_{1:t-1})$.

Once these filtering densities have been calculated and stored, we then simulate backwards. First $X_n$ is simulated from the filtering density $\Pr(X_n|y_{1:n})$; then for $t=n-1,\ldots,1$ we iteratively simulate $X_t$ given our simulated value for $X_{t+1}$, from
\[
 \Pr(X_t=l|y_{1:n},X_{t+1}=k)=\Pr(X_t=l|y_{1:t},X_{t+1}=k)\propto \Pr(X_t=l|y_{1:t})P_{lk}.
\]

The computational complexity of the forward-backward algorithm is $O(nK^2)$ for the forward recursion, and $O(nK)$ for the backward simulation. This compares with $O(nK)$ for applying the single-site update to all state-values. Thus, particularly for large $K$ and where the dependence in the state model is weak, 
single-site updates may be more efficient. 

In the above description, we supressed the dependence on the unknown parameter $\theta$. Standard MCMC algorithms will update $X_{1:n}$ given $\theta$ and then $\theta$ given $X_{1:n}$ in one iteration. Thus each iteration will (potentially) have a new $\theta$ value, and will require the re-application of the forward-backward algorithm to simulate $X_{1:n}$.
One approach to reducing the computational cost of using the forward-backward algorithm within MCMC, suggested by \citet{Fearnhead:2006SC} is to (i) obtain a good point estimate of the parameters, $\hat{\theta}$; (ii) apply the forward recursion 
for this value of the parameter; and (iii) at each iteration use $\Pr(X_{1:n}|y_{1:n},\hat{\theta})$ as an independence proposal for updating the state.
The advantage of this is that the costly forward-recursion is only required once, as opposed to at every iteration of the MCMC algorithm. Furthermore, \citet{Fearnhead:2006SC} describe an efficient algorithm for simulating large samples of $X_{1:n}$ from the backward simulation step. In applications, providing a good estimate is obtained in (i), this approach has shown to produce efficient MCMC updates. Estimation in (i) could be performed in an adaptive manner during the burn-in period of the MCMC algorithm.

Our forward-backward description has focussed on discrete-time processes. It is possible to extend the idea to continuous-time (though still discrete valued) HMMs. See for example \citet{Fearnhead/Meligkotsidou:2004} and \citet{Fearnhead/Sherlock:2006}.

\noindent{\bf Example 1: Block independence sampler}

For the SV model of Example 1, we cannot sample directly from the full conditional distribution $p(x_{1:n}|y_{1:n})$. Instead we follow \citet{Shephard/Pitt:1997} and consider an independence sampler for block updating. The proposal distribution for the independence sampler is based on a natural extension of the independence sampler for singe-site updates.

Consider an update for $X_{t:s}$ for $s>t$. For an efficient independence proposal we require a good approximation to
$p(x_{t:s}|x_{t-1},x_{s+1},y_{t:s})$. (If $t=1$ we would drop the conditioning on $x_{t-1}$, and if $s=n$ we would drop the conditioning on $x_{s+1}$ here and in the following.) Now we can write
\[
 p(x_{t:s}|x_{t-1},x_{s+1},y_{t:s}) \propto p(x_{t:s}|x_{t-1},x_{s+1})\prod_{j=t}^s p(y_j|x_j),
\]
where the first term on the right-hand side is a multivariate Gaussian density. Thus if for all $j=t,\ldots,s$, we approximate $p(y_j|x_j)$ by a Gaussian likelihood, we obtain a Gaussian approximation to $p(x_{t:s}|x_{t-1},x_{s+1},y_{t:s})$ which can be used as an independence proposal. We can obtain a Gaussian approximation to $p(y_j|x_j)$ by using a quadratic (in $x_j$) approximation to $\log p(y_j|x_j)$ via a Taylor expansion about a suitable estimate $\hat{x}_j$. The details of this quadratic approximation are the same as for the single-step update described above; see 
\citet{Shephard/Pitt:1997}. The resulting quadratic approximation to $p(x_{t:s}|x_{t-1},x_{s+1},y_{t:s})$ can be calculated efficiently using the Kalman Filter \citep{Kalman/Bucy:1961}, or efficient methods for Gaussian Markov Random Field models \citep{Rue/Held:2005}, and its complexity is $O(s-t)$.

Implementation of this method requires a suitable set of estimates $\hat{x}_{t:s}=(\hat{x}_t,\ldots,\hat{x}_s)$. If we denote by
$q(x_{t:s}|\hat{x}_{t:s})$ the Gaussian approximation to $p(x_{t:s}|x_{t-1},x_{s+1},y_{t:s})$ obtained by using the estimate 
$\hat{x}_{t:s}$, then one approach is to: (i) choose an initial estimate $\hat{x}_{t:s}^{(0)}$; and (ii) for $i=1,\ldots,I$ set $\hat{x}_{t:s}^{(i)}$ to be the mean of $q(x_{t:s}|\hat{x}_{t:s}^{(i-1)})$. In practice choosing $\hat{x}_{t:s}^{(0)}$ to be the mean of $p(x_{t:s}|x_{t-1},x_{s+1})$ and using small values of $I$ appears to work well.

This approach to designing independence proposals can be extended to other models where the model of the state is linear-Gaussian \citep[see][]{Jungbacker/Koopman:2007}. Using the resulting independence sampler within an MCMC algorithm is straightforward if it is efficient to update the complete state path $X_{1:n}$. If not, we must update the state in smaller blocks. A simplistic approach would be to split the data into blocks of (approximately) equal size, $\tau$ say, and then update in turn $X_{1:\tau}$, $X_{(\tau+1):2\tau}$ etc. However, this approach will mean that state values towards the boundaries of each block will mix slowly due to the conditioning on the state values immediately outside the boundary of the blocks. To avoid this \citet{Shephard/Pitt:1997} suggest randomly choosing the blocks to be updated for each application of the independence proposal. Another popular alternative is to choose overlapping blocks, for example $X_{1:2\tau}$, $X_{(\tau+1):3\tau}$, $X_{(2\tau+1):4\tau}$ and so on. 

A further important implementation consideration is the choice of block size. Too small and we obtain poor mixing due to the strong dependence of $X_{t:s}$ on $X_{t-1}$ and $X_{s+1}$; too large and we have poor mixing due to low acceptance rates. One approach is to use adaptive MCMC methods to choose appropriate block sizes, see \cite{Roberts/Rosenthal:2006}. Here we will look at the effect that block size has on acceptance probabilities for the SV model.  
\begin{figure}
\begin{center}
 \includegraphics[height=4in,width=5in,angle=0]{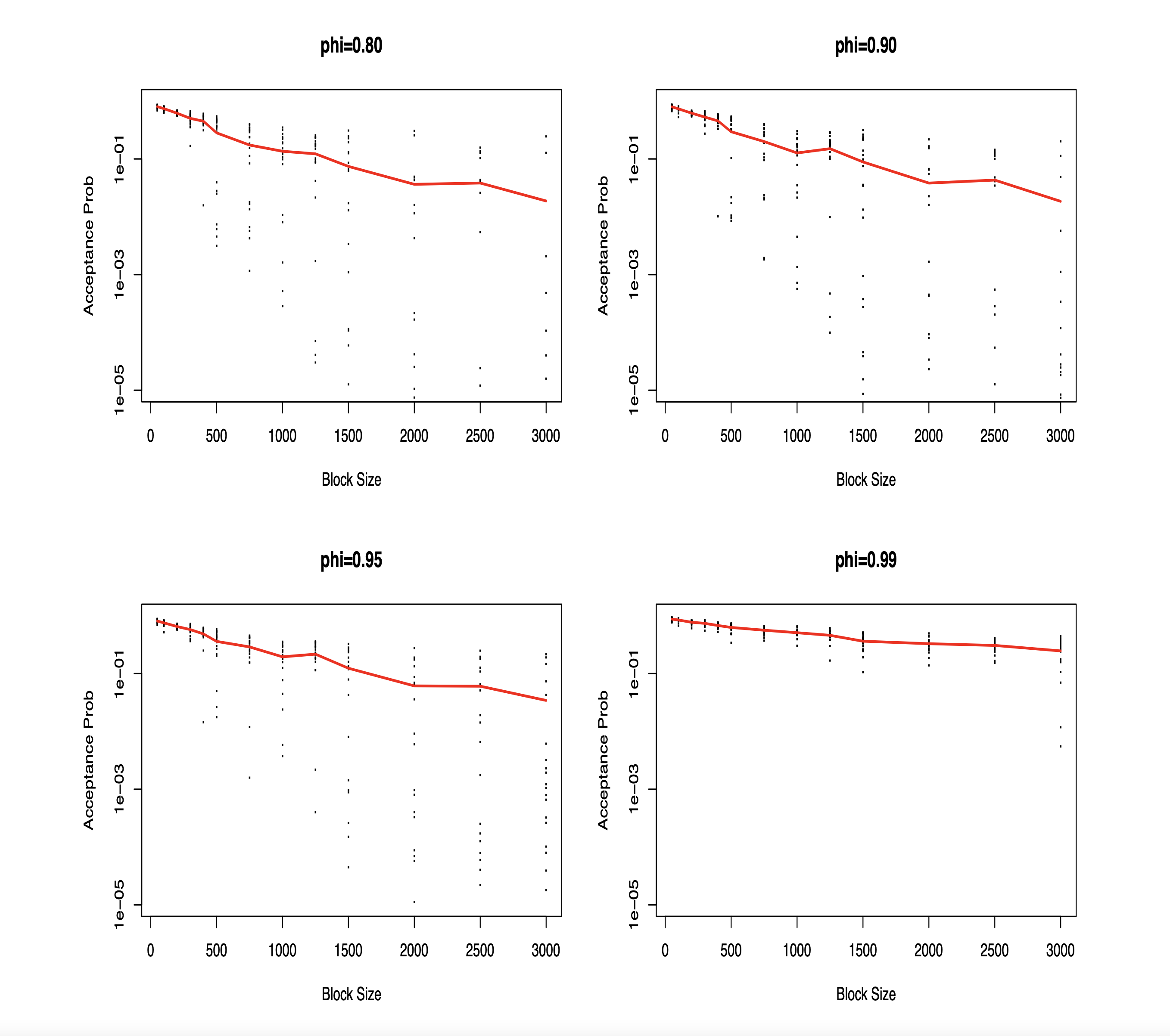}
 \caption{\label{fig:SVblock}Average acceptance rates for different block sizes, and different $\phi$ values. Black dots show mean acceptance rates for 20 different data-sets for each block size. Red lines show mean acceptance rates for each block size. All runs had $\tau^2=\sigma^2/(1-\phi^2)=0.2$. (Some MCMC runs had acceptance rates that are too small to appear on the plot.)}
\end{center}
\end{figure}

Plots of average acceptance rates for different block sizes and different data sets are shown in Figure \ref{fig:SVblock}. Two features are striking. The first is that efficiency varies substantially with $\phi$, with values of $\phi\approx1$ producing higher average acceptance rates. This is because for $\phi\approx1$ there is stronger dependence in the state-process, and thus the (Gaussian) $p(x_{t:s}|x_{t-1},x_{s+1})$ dominates the (non-Gaussian) likelihood $p(y_{t:s}|x_{t:s})$. The second is that there is great variability in acceptance rates across different runs: thus choice of too large block sizes can lead to the chain becoming easily stuck (for example, acceptance probabilities of $10^{-8}$ or less were observed for blocks of 2,000 or more observations when $\phi=0.8$). This variability suggests that either randomly choosing block sizes, or adaptively choosing block sizes for a given data set are both sensible strategies.

However, overall we see that the block updates are particularly efficient for the SV model. For block updates, acceptance rates $>0.01$ are reasonable, and the average acceptance rate was greater than this for all combinations of $\phi$ and block size that we considered. Even looking at the worst-case acceptance rates across all runs, we have acceptances rates greater than $0.01$ for blocks of size 400 when $\phi=0.8$; and for $2,500$ when $\phi=0.99$.




\section{Updating the parameters}

The natural approach to updating $\theta$ within the MCMC algorithm is to condition on the value of the state path $x_{1:n}$. Often this is straightforward as either conjugate priors for $\theta$ can be chosen so that we can sample directly from $p(\theta|x_{1:n},y_{1:n})$, or $\theta$ is of sufficiently low-dimension that we can use efficient independence proposals. Sometimes, 
we need to update components or blocks of $\theta$ at a time, rather than the updating the whole parameter vector in one go.

However, even if we can sample from the full-conditional $p(\theta|x_{1:n},y_{1:n})$, the overall efficiency of the MCMC algorithm can still be poor if there is strong correlation between $\theta$ and $x_{1:n}$. The rate of convergence of an algorithm that alternates between sampling from $p(x_{1:n}|\theta,y_{1:n})$ and $p(\theta|x_{1:n},y_{1:n})$ is given by \citet{Liu:1994} and \citet{Roberts/Sahu:1997}. If for a square-integrable function $f$ of the parameters, we define the {\em{Bayesian fraction of missing information}}:
\begin{equation} \label{eq:gf}
 \gamma_f=1-\frac{\mbox{E}\left(\mbox{Var}\left(f(\theta)|X_{1:n},y_{1:n} \right)|y_{1:n} \right)}{\mbox{Var}\left(f(\theta)|y_{1:n} \right)},
\end{equation}
then the geometric rate of convergence of the MCMC algorithm is $\gamma=\sup_{f}\gamma_f$. Values of $\gamma\approx1$ suggest a poorly mixing MCMC algorithm. This will occur when, after conditioning on the data, there are functions $f$ for which most of the variation in $f(\theta)$ is explained by the value of the state, $X_{1:n}$.

When there is strong dependence between $\theta$ and $X_{1:n}$, there are two techniques for improving mixing. The first is to consider a different parameterisation, with the hope that for this new parameterisation there will be less dependence between the state and the parameter. The second is to use moves that jointly update $\theta$ and $X_{1:n}$. We will describe and evaluate approaches for updating $\theta$ given $X_{1:n}$, and then consider these two approaches for improving mixing in turn.

\subsection{Conditional updates of the parameters}
\label{sec.PFCS.condupdates}

Here we focus on Examples 1 and 2, outlining how parameter updates can be made with these models, and investigating the mixing properties of the resulting MCMC algorithms.

\noindent{\bf Example 1: Conditional parameter updates}

Following \citet{Shephard/Pitt:1997} we will consider independent priors for $\beta$, $\sigma^2$ and $\phi$. As $\beta$ is a scale parameter, we choose the canonical uninformative prior, $p(\beta)\propto1/\beta$. For $\sigma^2$ our prior is $S_0\chi^{-2}_p$, where $X\sim \chi^{-2}_\nu \Leftrightarrow 1/X \sim \chi^2_\nu$. As it is normal to restrict $|\phi|<1$, we choose a $\mbox{Beta}(a,b)$ prior for $(\phi+1)/2$. For these choices we have that conditional on $\{x_{1:n},y_{1:n}\}$, $\beta$ is independent of $\phi,\sigma^2$, and has distribution 
\begin{equation} \label{eq:b2}
\beta^2|\{x_{1:n},y_{1:n}\}\sim \chi^{-2}_n \sum_{t=1}^ny_t^2\exp\{-x_t\}.
\end{equation}
To update $\phi$ and $\sigma$ it is simplest to use their conditional distributions
\[
 \sigma^2|\{x_{1:n},y_{1:n},\phi\}\sim\chi^{-2}_{n+p}\left\{S_0+x_1^2(1-\phi^2)+\sum_{t=2}^n(x_t-\phi x_{t-1})^2 \right\},
\]
\[
 p(\phi|x_{1:n},y_{1:n},\sigma) \propto (1+\phi)^{a-1/2}(1-\phi)^{b+1/2}\exp\left\{-\frac{(1-\phi^2)x_1^2}{2\sigma^2}-\frac{1}{2\sigma^2}\sum_{t=2}^n(x_t-\phi x_{t-1})^2 \right\}.
\]
The distribution for $\sigma^2$ can be sampled from directly. For $\phi$, a simple procedure is an independence sampler with Gaussian proposal. The Gaussian proposal is chosen proportional to
\[
\exp\left\{-\frac{(1-\phi^2)x_1^2}{2\sigma^2}-\frac{1}{2\sigma^2}\sum_{t=2}^n(x_t-\phi x_{t-1})^2 \right\}, 
\]
which corresponds to a mean of $\sum_{t=2}^nx_tx_{t-1}/\sum_{t=2}^{n-1}x_t^2$ and a variance of $\sigma^2/\sum_{t=2}^{n-1}x_t^2$. This distribution can propose values outside $(-1,1)$, and such values will always be rejected.

\begin{table}
  \caption{\label{Tab:ACFSV} Lag-1 autocorrelation for $\beta$ for both Non-centered and Centered parameterisations. Results are for $\sigma^2=0.02^2$, $\beta=1$ and $n=200$, and different values of $\phi$.}
\begin{center}
\begin{small}
\begin{tabular}{c|ccccc}
  \hline
 $\phi$ & 0.8 & 0.9 & 0.95 & 09.75 & 0.99  \\ \hline
Non-centered & 0.11 & 0.21 & 0.37 & 0.62 & 0.98\\
Centered & 0.89 & 0.79 & 0.64 & 0.43 & 0.29 \\
\hline
\end{tabular}
\end{small}
\end{center}
\end{table}

An example of how the mixing of the MCMC algorithm is affected by the dependence within the state-model is shown in the top row of Table \ref{Tab:ACFSV} (labelled non-centered parameterisation). As $\phi$ increases, that is the dependence in the state model increases, then the mixing deteriorates. This is because in this limit the amount of additional information about $\beta$ from observing the state-path (in addition to the observations) remains roughly constant as $\phi$ increases, but the amount of information about $\beta$ contained just in the observations is decreasing. This means that the Bayesian fraction of missing information is increasing, and thus the MCMC algorithm mixes more poorly.

\noindent{\bf Example 2: Conditional parameter updates}

Let $P_k$ denote the $k$th row of the transition matrix, $P$. Furthermore consider the case where the parameter vector can be written as $\theta=(P,\phi_1,\ldots,\phi_K)$, with the likelihood function given $X_t=k$ being of the form $f_k(y|\theta)=f_k(y|\phi_k)$. That is we have a disjoint set of parameters for each of the $K$ likelihood models. Further assume first that the distribution of $X_1$ is independent of $\theta$. In this case, if our priors for the $P_k$s and $\phi_k$s are independent, then the full conditional $p(\theta|x_{1:n},y_{1:n})$ simplifies. Conditional on $\{x_{1:n},y_{1:n}\}$, we have independence of $P_1,\ldots,P_K,\phi_1,\ldots,\phi_K$. Thus we can perform independent updates on each of these $2K$ sets of parameter in turn.
(If the distribution of $X_1$ depends on $P$, then this will introduce weak dependence in the posterior distribution of the $P_k$s.)

 If we choose a Dirichlet prior for the entries of $P_k$, then each  $p(P_k|x_{1:n},y_{1:n})$ will be  a Dirichlet distribution.
Updating of $\phi_k$ will depend on the specific likelihood model and priors used. However, for the DNA model introduced in Section \ref{SSUP}, we have $\phi_k=\pi^{(k)}=(\pi^{(k)}_{\mbox{\small{A}}},\pi^{(k)}_{\mbox{\small{C}}},\pi^{(k)}_{\mbox{\small{G}}},\pi^{(k)}_{\mbox{\small{T}}})$, and if we have a Dirichlet prior then  $p(\phi_k|x_{1:n},y_{1:n})$ will again be Dirichlet.



\subsection{Reparameterisation of the model}

We have seen that dependence between $X_{1:n}$ and $\theta$ can result in a MCMC algorithm for $(X_{1:n},\theta)$ that mixes poorly. One approach to alleviate this is to consider alternative parameterisations.

\citet{Papaspiliopoulos/Roberts/Skold:2007} describe two possible general parameterisations for hierarchical models \citep[see also][]{Gelfand/Sahu/Carlin:1995,Papaspiliopoulos/Roberts/Skold:2003}, and these can be used for state-space models. These are {\em{centered parameterisations}}, which in our set-up is defined by a model where $p(\theta|x_{1:n},y_{1:n})=p(\theta|x_{1:n})$, and {\em{non-centered parameterisations}} where a priori $\theta$ and $X_{1:n}$ are independent. For the stochastic volatility model of Example 1 our parameterisation for $\beta$ is  non-centered since our model for $X_{1:n}$ does not depend on $\beta$. In contrast, both $\sigma$ and $\phi$ are centred parameterisations, as is $P$ in Example 2.

While it is non-trivial to introduce a non-centered parameterisation for Example 2 \citep[though][propose approaches that could be used]{Papaspiliopoulos:2003,Roberts/Papaspiliopoulos/Dellaportas:2004}, it is straightforward to introduce a centered parameterisation for $\beta$ in Example 1.  We define $\mu=2\log\beta$ and a new state model $X'_{1:n}$ where
\[
X'_t|\{x'_{1:t-1},y_{1:t-1}\}\sim\mbox{N}(\mu+\phi(x'_{t-1}-\mu),\sigma^2),
\]
with $X'_1\sim\mbox{N}(\mu,\sigma^2/(1-\phi^2))$, and
\[
Y_t|\{x'_{1:t},y_{1:t-1}\}\sim\mbox{N}(0,\exp\{x'_t\}).
\]

For this parameterisation we have \citep{Pitt/Shephard:1999}
\[
\mu|\{x'_{1:n},y_{1:n}\} \sim \mbox{N}(b/a,\sigma^2/a),
\]
where $a=(n-1)(1-\phi)^2+(1-\phi^2)$ and $b=(1-\phi)\{\sum_{t=2}^n(x'_t-\phi x'_{t-1})\}+x'_1(1-\phi^2)$.

For large $n$ we can compare $\gamma_f$ (\ref{eq:gf}) for $f(\theta)=\mu$ for both centered and non-centered parameterisations.
If we conjecture that $\gamma\approx\gamma_f$, then these values will inform us about the relative efficiency of the two parameterisations. To compare $\gamma_f$ for the two parameterisations we need only compare
$\mbox{E}(\mbox{Var}(2\log\beta|X_{1:n},y_{1:n})|y_{1:n})$ and $\mbox{E}(\mbox{Var}(\mu|X'_{1:n},y_{1:n})|y_{1:n})$. If the former is larger, then the non-centered parameterisation will have a smaller value for $\gamma_f$, and, we may conjecture, will have a better rate of convergence. Otherwise $\gamma_f$ will be smaller for the centered parameterisation.

Now, for the centered parameterisation we have $\mbox{Var}(\mu|X'_{1:n},y_{1:n})=\sigma^2/a\approx \sigma^2/(n(1-\phi)^2)$. Thus, as
this does not depend on $X'_{1:n}$, we have
\[
\mbox{E}(\mbox{Var}(\mu|X'_{1:n},y_{1:n})|y_{1:n})\approx\frac{\sigma^2}{n(1-\phi)^2}.
\]
For the non-centered parameterisation, from (\ref{eq:b2}), we have that $\mbox{E}(\mbox{Var}(2\log\beta|X_{1:n},y_{1:n})|y_{1:n})=\mbox{Var}(\log \chi^2_n)$, thus for large $n$
\[
\mbox{E}(\mbox{Var}(2\log\beta|X_{1:n},y_{1:n})|y_{1:n})\approx \frac{2}{n}.
\]
Thus $\gamma_f$ is smaller for the centered parameterisation if $2/n<\sigma^2/(n(1-\phi)^2)$ or
\[
\phi>1-\frac{\sigma}{\sqrt{2}}.
\]
This suggests that as $\phi\rightarrow1$ we should prefer using the centered parameterisation, but for small $\phi$ the non-centered parameterisation would be preferred. This is confirmed by simulation (see Table \ref{Tab:ACFSV}). Similarly, when $\sigma$ is small we should prefer the centered parameterisation. 

For the specific model we consider in Example 1, we have centered parameterisations for $\sigma$ and $\phi$. It is possible to extend the non-centered parameterisations for $\beta$ to one for $(\beta,\sigma)$ and even $(\beta,\sigma,\phi)$. For $(\beta,\sigma)$ we introduce a state $X'_{1:n}$ where
\[
X'_t|\{x'_{1:t-1},y_{1:t-1}\}\sim\mbox{N}(\phi x'_{t-1},1),
\]
with $X'_1\sim\mbox{N}(0,1/(1-\phi^2))$, and
\[
Y_t|\{x'_{1:t},y_{1:t-1}\}\sim\mbox{N}(0,\beta^2\exp\{\sigma x'_t\}).
\]
For $(\beta,\sigma,\phi)$ we can parameterise the state in terms of the standardised residuals in the AR model, $(X_t-\phi X_{t-1})/\sigma$, and $X_1\sqrt{1-\phi^2}$, which are independent standard normal random variables. This latter idea, and ideas related to it, has been used extensively within continuous-time stochastic volatility models \citep[see e.g.][]{Roberts/Stramer:2001,golightly2008bayesian}.

\subsection{Joint updates of the parameters and state}
One way of thinking about why strong correlation between $\theta$ and $X_{1:n}$ produces poor mixing, is that large moves of $\theta$ are likely to be rejected as they will be inconsistent with the current value of the state. This will happen even if the proposed new value for $\theta$ is consistent with the data. This motivates jointly updating $\theta$ and $X_{1:n}$, from a proposal
$q(\theta',x'_{1:n}|\theta,x_{1:n})=q(\theta'|\theta)q(x'_{1:n}|\theta')$. Thus $q(\theta'|\theta)$ could propose large moves, and then values of the state-process consistent with $\theta'$ will be simulated from $q(x'_{1:n}|\theta')$.

This is most easily and commonly implemented for models where we can simulate directly from $p(x_{1:n}|\theta,y_{1:n})$, in which case
we choose $q(x'_{1:n}|\theta')=p(x'_{1:n}|\theta',y_{1:n})$. The resulting acceptance ratio then simplifies to:
\[
 \min\left\{1, \frac{q(\theta|\theta')p(\theta'|y_{1:n})}{q(\theta'|\theta)p(\theta|y_{1:n})}  \right\}.
\]
This acceptance ratio does not depend on $x_{1:n}$ or $x'_{1:n}$. The marginal chain for $\theta$ is equivalent to an MCMC chain for $p(\theta|y_{1:n})$ with proposal distribution $q(\theta'|\theta)$.

Providing an efficient proposal $q(\theta'|\theta)$ can be found, such an MCMC algorithm will always be more efficient than one that
updates $\theta$ and $X_{1:n}$ independently. However, the difficulty with implementing this idea is how to choose $q(\theta'|\theta)$.
For Markov-modulated Poisson processes, \citet{Sherlock/Fearnhead:2008} found that a Gibbs sampler that updated $X_{1:n}$ given $\theta$ and $\theta$ given $X_{1:n}$ performed better than this joint update where $q(\theta'|\theta)$ was chosen to be a symmetric random-walk. A further advantage of the Gibbs sampler, is that it avoids tuning $q(\theta'|\theta)$, though this problem can be alleviated by using adaptive MCMC schemes \citep{Sherlock/Fearnhead:2008,Andrieu/Thoms:2008}.

A simple extension of this joint updating idea is possible if we have an efficient independence proposal for $x_{1:n}$ given $\theta$ -- as this proposal could be used as $q(x'_{1:n}|\theta')$. Here the efficiency of the resulting algorithm will depend on both the efficiency of $q(\theta'|\theta)$ as a proposal for an MCMC that explores $p(\theta'|y_{1:n})$, and also the closeness of $q(x'_{1:n}|\theta')$ to $p(x'_{1:n}|\theta',y_{1:n})$. 

\section{Particle MCMC}

When interest lies in the parameters, rather than the hidden states, the most natural way to avoid the poor mixing that arises from the dependence between $X_{1:T}$ and $\theta$ is to perform inference on the marginal $\pi(\theta)\propto \pi_0(\theta) p(y_{1:T}|\theta)$. In general, of course, it is impossible to evaluate $p(y_{1:T}|\theta)=\int p(y_{1:T}|x_{1:T},\theta)p(x_{1:T}|\theta)\PFCSmd \theta$, and so this approach is not viable.

In the spirit of the above, however, the \emph{pseudo-marginal Metropolis-Hastings algorithm} of \cite{AndRob2009} replaces the true, unknown marginal with a non-negative, unbiased stochastic estimate of it. Provided the current estimate is carried forward to the next iteration along with the current state, quite remarkably, draws of $\theta$ from this algorithm, once it has converged, are draws from the true posterior, $\pi(\theta)$. 

\citet{AndDouHol2010} considered the case of inference for state space models where it is natural to obtain the estimate from a particle filter, leading to the particle Metropolis-Hastings algorithm. The output from the particle filter can also be used to obtain draws from $\pi(\theta,x_{1:T}|y_{1:T})$, if that is desired, leading to the \emph{particle marginal Metropolis-Hastings} (PMMH) algorithm. Alternatively, the \emph{particle Gibbs} algorithm allows sampling from the whole path, $x_{1:T}|\theta,y_{1:T}$; any valid Metropolis-Hastings or Gibbs move may then be used for $\theta|x_{1:T},y$. 

The particle filter is, essentially, a sequential importance sampler with one important tweak, so we commence with a description of importance sampling in the context of Bayesian statistics, before moving on to  the particle filter. This sets the groundwork for descriptions of pseudo-marginal Metropolis-Hastings, and the PMMH and particle Gibbs algorithms.

\subsection{Importance sampling}
\label{sec.ImpSamp}
Consider a general specification of a model, $p(x|\theta)$, for a latent state, $x$, given a parameter, $\theta$, and a second model for an observation, $y$, $p(y|x,\theta)$. This includes the full hidden Markov model, with $ x=x_{1:T}$, and $y=y_{1:T}$,
\begin{equation}
\label{eqn.naiveIS}
   p(x|\theta)=p(x_1|\theta)\prod_{t=2}^Tp(x_t|x_{t-1},\theta)
    ~~~\mbox{and}~~~
    p(y|x,\theta)=\prod_{t=1}^T p(y_t|x_t,\theta).
\end{equation} 
Setting $T=1$ in the above, we see that it also includes the model for the very first latent state, $x_1$, and the first observation, $y_1$. 

We might be interested in the likelihood, $p(y|\theta)=\int p(y,x|\theta)\PFCSmd x$ and, for any function $h(x)$, in $\PFCSExpect{h(X)}$ under the conditional distribution corresponding to $p(x|y,\theta)=p(y,x|\theta)/p(y|\theta)$, but both integrals are intractable.

Let $q$ be a density whose support includes that of  $p(x|\theta)$, but from which we can sample. Then
\[
\int h(x) p(y,x|\theta)\PFCSmd x
=
\int h(x) \frac{p(y,x|\theta)}{q(x)}~q(x)\PFCSmd x
=
\PFCSExpects{q}{h(X) w(X)},
\]
where $w(x)=p(y,x|\theta)/q(x)$. Setting $h(x)=1$, we obtain $
p(y|\theta)=\PFCSExpects{q}{w(X)}$,
and, hence,
\[
\PFCSExpects{p(x|y,\theta)}{h(X)}
=
\int h(x) \frac{p(y,x|\theta)}{p(y|\theta)}\PFCSmd x
=
\frac{\PFCSExpects{q}{h(X)w(X)}}{\PFCSExpects{q}{w(X)}}.
\]
This rearrangement still leaves us with two intractable integrals; however, we can approximate each of the expectations through Monte Carlo, by taking $M$ independent samples, $x^{1},\dots,x^{M}$ from $q$. By the strong law of large numbers, as $M\to \infty$, with probability $1$,
\begin{equation}
\label{eqn.Lconsistent}
\PFCSphat(y|\theta;x^{1:M}):=\frac{1}{M}\sum_{m=1}^M w\left(x^{m}\right)
\to
\PFCSExpects{q}{w(X)}= p(y|\theta)
\end{equation}
and, provided $\PFCSExpects{q}{|h(X)|w(X)}<\infty$,
$\frac{1}{M}\sum_{m=1}^M h\left(x^{m}\right)w\left(x^{m}\right)
\to
\PFCSExpects{q}{h(X)w(X)}$.
Hence
\[
\frac{\sum_{m=1}^M h\left(x^{m}\right)w\left(x^{m}\right)}
{\sum_{m=1}^M w\left(x^{m}\right)}
\to 
\PFCSExpects{p(x|y,\theta)}{h(X)}.
\]
Defining the normalised weight, $\PFCSwtil^{i}= w\left(x^{i}\right)/\sum_{m=1}^M w\left(x^{m}\right)$, we may write the above as $\sum_{m=1}^M h\left(x^{m}\right)\PFCSwtil^{m}
\to 
\PFCSExpects{p(x|y,\theta)}{h(X)}$. Since this holds for any $h$, the weighted sample $(x^{1},\PFCSwtil^{1})$ $,\dots,(x^{M},\PFCSwtil^M)$ can be viewed as a discrete approximation, $\PFCSphat(x|y,\theta;x^{1:M})$, to the distribution corresponding to $p(x|y,\theta)$. A measure of the accuracy of $\PFCSphat(x|y,\theta;x^{1:M})$ is the \emph{effective sample size} \cite[]{kong1994sequential},
\[
\mathsf{ESS}:=\frac{\left\{\sum_{m=1}^Mw\left(x^{m}\right)\right\}^2}{\sum_{m=1}^M\left\{w\left(x^{m}\right)\right\}^2}
=
\frac{1}{\sum_{m=1}^M\left\{\PFCSwtil^{m}\right\}^2},
\]
an approximate measure of the equivalent number of iid samples that would give the same variance for the approximation to $\PFCSExpects{p(x|y,\theta)}{h(X)}$. When all the weights are equal, the effective sample size is $M$ whereas when one of the weights is much larger than all of the others combined, it is close to $1$. 

In terms of ESS, the ideal proposal would be $q(x)=p(x|y,\theta)$, since then $\PFCSwtil^{m}=1/M$ for each $m$. In practice, to obtain a high effective sample size for a given $M$, we require $q$ to be ``close to" $p(x|y,\theta)$. Typically, the higher the dimension of $x$, the more difficult it is to find a tractable $q$ that is sufficiently close to $p(x|y,\theta)$, with the result that, in high-dimensional settings, most of the $\PFCSwtil$ are likely to be close to zero with only a few (or even just one) of the sampled points having non-negligible weight. To obtain a reasonable ESS, $M$ must, typically, increase exponentially with the dimension of $X$ so that importance sampling usually becomes infeasible for dimensions larger than about ten; this exponential increase is often referred to as \emph{the curse of dimensionality}.  

In Section \ref{sec.pseudo}, the unbiasedness of the estimator of the likelihood will be a crucial part of the argument for the validity of pseudo-marginal MCMC. Here, as required, 
\begin{equation}
\label{eqn.Lunbiased}
\PFCSExpects{q}{\PFCSphat(y|\theta;X^{1:M})}
=
\frac{1}{M}\sum_{m=1}^M
\PFCSExpects{q}{w(X^m)}
=
\PFCSExpects{q}{w\left(X^{1}\right)}
=
p(y|\theta).
\end{equation} 

\subsection{The Particle filter}
\label{sec.ParticleFilter}
Returning to the state-space model, we first consider inference for $x_{T}$ conditional on $y_{1:T}$ when $\theta$ is fixed and known, which leads to the particle filter. Since the $x_t$ process is Markovian, if we were to use importance sampling, it would be natural to consider a proposal of the form
\begin{equation}
\label{eq.jointProposal}
q(x_{1:T})=q(x_1)\prod_{t=2}^Tq(x_t|x_{t-1}).
\end{equation}
 Given the curse of dimensionality introduced in Section \ref{sec.ImpSamp}, the most obvious use of importance sampling, for $x_{1:T}$ \eqref{eqn.naiveIS}, is infeasible even for time series of moderate length as it is typically not possible to find a tractable $q(x_{1:T})$ that is even moderately close to $p(x_{1:T}|y_{1:T},\theta)$. However, in many scenarios, the state space itself is of low dimension (for example, a position in space or the logged change in price of a stock), so it \emph{is} feasible to use importance sampling for  $x_1|y_1,\theta$ as follows:

 \begin{itemize}
  \item[\textbf{Step 0}:] use the importance-sampling proposal $q_1(x_1)$ to obtain a discrete approximation to $p(x_1|y_1,\theta)$, via the weighted sample $\left(x_1^{1},w_1^{1}\right),\dots,\left(x_1^{M},w_1^{M}\right)$, where 
 \begin{equation}
 \label{eqn.wOne}
 w_1^{m}\equiv w_1(x_1^{m})
 =
\frac{p(y_1|x_1^{m},\theta)p(x_1^{m}|\theta)}{q(x_1^{m})}
 .
 \end{equation} 
 \end{itemize}
 We might then sample from $q(x_2|x_1)$, which approximates $p(x_2|x_1,y_{1:2},\theta)$, once for each of $x_1^{1},\dots,x_1^{M}$ to extend the samples to $x_{1:2}^{1},\dots,x_{1:2}^{M}$ and obtain new weights $w_2^{1},\dots,w_2^{M}$. We could then iterate this procedure to obtain samples $(x_{1:T}^{1},w(x_{1:T}^{1})),\dots,(x_{1:T}^{M},w(x_{1:T}^{M}))$. The $m$th weight would be 
 \[
\frac{ p(y_1,x_1^{m}|\theta)\prod_{t=2}^T p(y_t,x_t^{m}|y_{1:t-1},x_{1:t-1}^{m},\theta)}
{q(x_1^{m})\prod_{t=2}^Tq(x_t^{m}|x_{t-1}^{m})}
=
\prod_{t=1}^T w_t^{m},
\]
where $w_1^{m}$ is as defined in \eqref{eqn.wOne}  and, because of the special structure of the model,
\begin{equation}
\label{eqn.tWeight}
w_t^{m}
=
\frac{p(y_t,x_t^{m}|y_{1:t-1},x_{1:t-1}^{m},\theta)}
{q(x_t^{m}|x_{t-1}^{m})}
=
\frac{p(y_t|x_t^{m},\theta)p(x_{t}^{m}|x_{t-1}^{m},\theta)}
{q(x_t^{m}|x_{t-1}^{m})}~~(t\ge 2).
 \end{equation}
 The above sequential importance sampling algorithm is, unfortunately, simply a rewriting of the single-proposal importance sampler for $x_{1:T}$ that the curse of dimensionality has already forced us to dismiss. However, with a small tweak, an algorithm using sequential proposals can work remarkably well. 
 
 The sequential importance sampler fails because even though the individual weights $w_t^{m}$ might be well behaved, typically, the product of $T$ such weights is not. The particle filter circumvents this issue by \emph{resampling} from the weighted sample $(x_{1}^{1},w_{1}^{1}),\dots,(x_{1}^{M},w_{1}^{M})$ to obtain an \emph{equally weighted} sample,  $\PFCSxtil_1^{1:M}$, which is also an approximation to $p(x_1|y_1,\theta)$. 
 
 The same resampling idea  applied to a weighted sample from $p(x_{t-1}|y_{1:t-1},\theta)$ produces an unweighted sample from the same distribution. From an unweighted sample, $\PFCSxtil_{t-1}^{1:M}$ ($t=2,\dots,T$), for each $m$, we then propose $x_t^{m}$ from $q(x_t|\PFCSxtil_{t-1
}^{m})$ and find $w_t^{m}$ according to \eqref{eqn.tWeight} but replacing $x_{t-1}^{m}$ with $\PFCSxtil_{t-1}^{m}$. 

Thus, a particle filter obtains a weighted sample approximation $(x_{1}^{1},w_{1}^{1}),\dots,(x_{1}^{M},w_{1}^{M})$ to $p(x_1|y_1,\theta)$ via importance sampling using Step 0. Then, given a weighted sample from $p(x_{t-1}|y_{1:t-1},\theta)$, it  repeats the following steps to obtain a weighted sample from $p(x_{t}|y_{1:t},\theta)$:
\begin{itemize}
\setlength\itemsep{0ex}
    \item[\textbf{Resample}:] Sample $M$ times from the weighted approximation $\left(x_{t-1}^{1},w_{t-1}^{1}\right),\dots,\left(x_{t-1}^{M},w_{t-1}^{M}\right)$ to $p(x_{t-1}|y_{1:t-1},\theta)$ to obtain an unweighted approximation $\PFCSxtil_{t-1}^{1:M}$ to the same distribution.
    \item[\textbf{Propagate}:] For each $m=1,\dots,M$, sample $x_t^{m}$ from a proposal $q(x_t|\PFCSxtil_{t-1}^{m})$.
    \item[\textbf{Weight}:] For each $m=1,\dots,M$, set 
    $
w_t^{m}
=
{p(y_t|x_t^{m},\theta)p(x_{t}^{m}|\PFCSxtil_{t-1}^{m},\theta)}/
{q(x_t^{m}|\PFCSxtil_{t-1}^{m})}
    $.
\end{itemize}
The $M$ samples are called \emph{particles}. 
At any time, $t$, the output from the particle filter is the weighted-sample approximation to $p(x_t|y_{1:t},\theta)$. 

The most commonly used special case of the above algorithm is the original, \emph{Bootstrap particle filter} of \cite{GSS1993}, which sets $q(x_1)=p(x_1)$ and $q(x_t|x_{t-1})=p(x_t|x_{t-1})$. In this case the $m$th weight simplifies to $w_t^{m}=p(y_t|x_t^{m})$. Importantly, the Bootstrap particle filter can be applied in situations where we can sample from $p(x_t|x_{t-1})$ even if we cannot analytically calculate this transition density; for example when the hidden process is an intractable diffusion that we can simulate from to desired accuracy via  a suitably fine Euler-Maruyama discretisation. Numerous extensions and enhancements to the particle filter presented here have been created; see \cite{chopin2020introduction} for further details. 

The key ingredient in the particle Metropolis-Hastings algorithm is an additional, optional, particle-filter output: an estimate of the likelihood. Using \eqref{eqn.Lconsistent}, the mean of the importance sample weights at time 1, provides a consistent estimator of $p(y_1)$. 
Similarly, the mean of the importance sample weights at time $t$ is a natural estimator of $p(y_t|y_{1:t-1})$.

We can, thus, obtain an estimator for the likelihood by the product of the average of the importance sampling weights at each time point. It will be helpful to write this estimator as a function of $u$, the realisation of $U$, the random variables used in the particle filter:
\begin{equation}
\label{eqn.PFphat}
\PFCSphat(y_{1:T}|\theta;u)=\prod_{t=1}^T\frac{1}{M}\sum_{m=1}^Mw_t^{m}.
\end{equation}
Just as the likelihood estimator obtained from importance sampling \eqref{eqn.Lconsistent} is unbiased \eqref{eqn.Lunbiased}, so is this estimator obtained from the particle filter \eqref{eqn.PFphat}. The proof is, however, less straightforward; see, \eqref{eqn.prove.unbiased} or, for example, \citet{PitSilGioKoh2012}.
 
\subsection{Pseudo-marginal Metropolis-Hastings}
\label{sec.pseudo}
We have seen that both importance sampling and the particle filter provide an unbiased estimator of the likelihood. Since the weights are all non-negative, the estimators are non-negative, too.

In general, from a non-negative, unbiased estimator of the likelihood $\PFCSphat(y_{1:T}|\theta;U)$ and a prior $\pi_0(\theta)$ we obtain a non-negative unbiased (up to a fixed, multiplicative constant) estimator of the posterior:
\begin{equation}
\label{eqn.ubLikeToPost}
\PFCSpihat(\theta;U)=\pi_0(\theta) \PFCSphat(y_{1:T}|\theta;U),
\end{equation}
where the random variables $U$ have arisen from some distribution represented by $q_*(u|\theta)$.

Given a current parameter value $\theta_i$ and a realisation of the unbiased estimator of the posterior $\PFCSpihat(\theta_i;u_i)$, the pseudo-marginal Metropolis-Hastings algorithm \citep[]{AndRob2009} proceeds as follows:
\begin{enumerate}
\setlength\itemsep{0ex}
\item Propose a new parameter value $\theta'$ from some $q(\theta'|\theta_i)$.
\item Sample $u'$ from $q_*(u'|\theta')$ to obtain $\PFCSpihat(\theta';u')$ via \eqref{eqn.ubLikeToPost}.
\item Calculate the Metropolis-Hastings acceptance probability:
\[\alpha_{pm}(\theta_i,u_i;\theta',u')
=
1\wedge \frac{\PFCSpihat(\theta';u')q(\theta_i|\theta')}{\PFCSpihat(\theta_i;u_i)q(\theta'|\theta_i)}.
\]
\item With a probability of $\alpha_{pm}(\theta_i,u_i;\theta',u')$ set $\theta_{i+1}\leftarrow \theta'$ and $u_{i+1}\leftarrow u'$ so that $\PFCSpihat(\theta_{i+1};u_{i+1})=\PFCSpihat(\theta';u')$; otherwise $\theta_{i+1}\leftarrow \theta_i$ and $u_{i+1}\leftarrow u_i$ so that $\PFCSpihat(\theta_{i+1};u_{i+1})=\PFCSpihat(\theta_i;u_i)$.
\end{enumerate}

The pseudo-marginal Metropolis-Hastings algorithm satisfies detailed balance with respect to $\widetilde{\pi}(\theta,u) \propto q_*(u|\theta)\PFCSpihat(\theta;u)$ since
\begin{align*}
    \widetilde{\pi}(\theta,u)~q(\theta'|\theta)q_*(u'|\theta') ~\alpha_{pm}(\theta,u;\theta',u')
    &=
    q_*(u|\theta)q_*(u'|\theta')
    ~
    \left\{\PFCSpihat(\theta;u)~q(\theta'|\theta)\wedge \PFCSpihat(\theta';u')~q(\theta|\theta')\right\}\\
    &=
    \widetilde{\pi}(\theta',u')~q(\theta|\theta')q_*(u|\theta) ~\alpha_{pm}(\theta',u';\theta,u).
\end{align*}
The marginal for $\theta$ is, therefore, proportional to
\[
\int q_*(u|\theta) \PFCSpihat(\theta;u) \PFCSmd u=\PFCSExpects{q_*}{\PFCSpihat(\theta;U)}
\propto 
\pi(\theta),
\]
since the estimator is unbiased. Hence, running the pseudo-marginal Metropolis-Hastings algorithm to convergence and discarding the auxiliary variables, $u$, provides a sample from the true posterior for $\theta$. Inference for the parameters of a state space model may, therefore, be performed by using likelihood estimates obtained by running a particle filter. This is the essence of the particle Metropolis-Hastings algorithm.

Pseudo-marginal Metropolis-Hastings mimics the idealised true marginal algorithm that would only be implementable if the true likelihood $p(y_{1:T}|\theta)$ were available. Its acceptance rate at stationarity is always below that of the idealised algorithm, since the former is
\begin{align*}
    \int \widetilde{\pi}(\theta,u) q(\theta'|\theta) q_*(u'|\theta') &\alpha_{pm}(\theta,u;\theta',u')  \PFCSmd u \PFCSmd u' \PFCSmd \theta \PFCSmd \theta'\\
    &=\int \PFCSExpects{q_*(u|\theta)q_*(u'|\theta')}{\PFCSpihat(\theta;U)q(\theta'|\theta)\wedge \PFCSpihat(\theta';U')q(\theta|\theta')}\PFCSmd \theta \PFCSmd \theta'\\
    &\le \int \pi(\theta)q(\theta'|\theta)\wedge \pi(\theta')q(\theta|\theta')\PFCSmd \theta \PFCSmd \theta'\\
    &= \int \pi(\theta) q(\theta'|\theta) \left\{1\wedge \frac{\pi(\theta')q(\theta|\theta')}{\pi(\theta)q(\theta'|\theta)}\right\}\PFCSmd \theta \PFCSmd \theta',
\end{align*}
which is the latter. Here we have applied Jensen's inequality twice to the concave function $h(x)=1\wedge x=x\wedge 1$. Noting this, \citet{AndVih2015} proves that a pseudo-marginal MH algorithm can never be as efficient as the ideal marginal MH algorithm. 

In Step 4 of the pseudo-marginal MH algorithm, when a proposal is accepted, so is the estimate of the posterior that is associated with it,  $\PFCSpihat(\theta';u')$. This then appears in the denominator of the acceptance ratio at Step 3 of the algorithm in all subsequent iterations until a new proposal is accepted. If $\PFCSpihat(\theta';u')$ happens to be a particularly large overestimate then subsequent acceptance ratios may be particularly small and it may take many iterations until a new proposal is accepted. This phenomenon is often referred to as the \emph{stickiness} of the pseudo-marginal Metropolis-Hastings algorithm. If the weights $w_t^m$ are bounded above then the potential deterioration of the algorithm is also bounded; otherwise it is not. 

When an estimate of the likelihood is consistent as the number of samples or particles $M\to \infty$, choosing a large $M$ decreases the chance and extent of poor behaviour. At the same time, the computational cost per iteration is, typically, proportional to $M$, so that in a fixed amount of CPU time, the number of iterations decreases. For likelihood estimates arising from a particle filter, the balance between the choice of the number of particles and the number of effective samples per second from the pseudo-marginal  Metropolis-Hastings algorithm was investigated simultaneously in \cite{DouPitDelKoh2015} and \cite{SheThiRobRos2015} using different methods but with similar resulting advice: pick a representative parameter value, $\widetilde{\theta}$, and choose the number of particles, $M$, such that $\mathsf{Var}[\log \PFCSphat(\widetilde{\theta}|y_{1:T};U)]\approx 1$. In contrast, if the estimate arises from importance sampling then unless it is possible to utilise multiple CPU cores it is usually better to use a single sample \citep[]{SheThiLee2017}.

\noindent{\bf Example 1: Pseudo-marginal Metropolis-Hastings}

We generated $T=400$ data points using $(\beta,\phi,\sigma)=(1,0.98,0.2)$, and set the priors as in Section \ref{sec.PFCS.condupdates}, with $a=b=1$, $\nu=5$ and $\nu/S_0=1/0.2^2$. Inference was conducted on $\theta:=(\log \beta,\mathsf{logit}(1/2+\phi/2), \log \sigma)$, with the priors transformed via the appropriate Jacobian term. Random walk Metropolis proposals $\theta'\sim \mathsf{N}(\theta,\lambda^2 \widehat{V})$ were employed, with $\lambda=1.3$ and $\widehat{V}$ an approximate posterior variance matrix obtained from an initial tuning run.

\begin{figure}
\begin{center}
 \rotatebox{90}
{\includegraphics[height=5in,width=5in,angle=270]{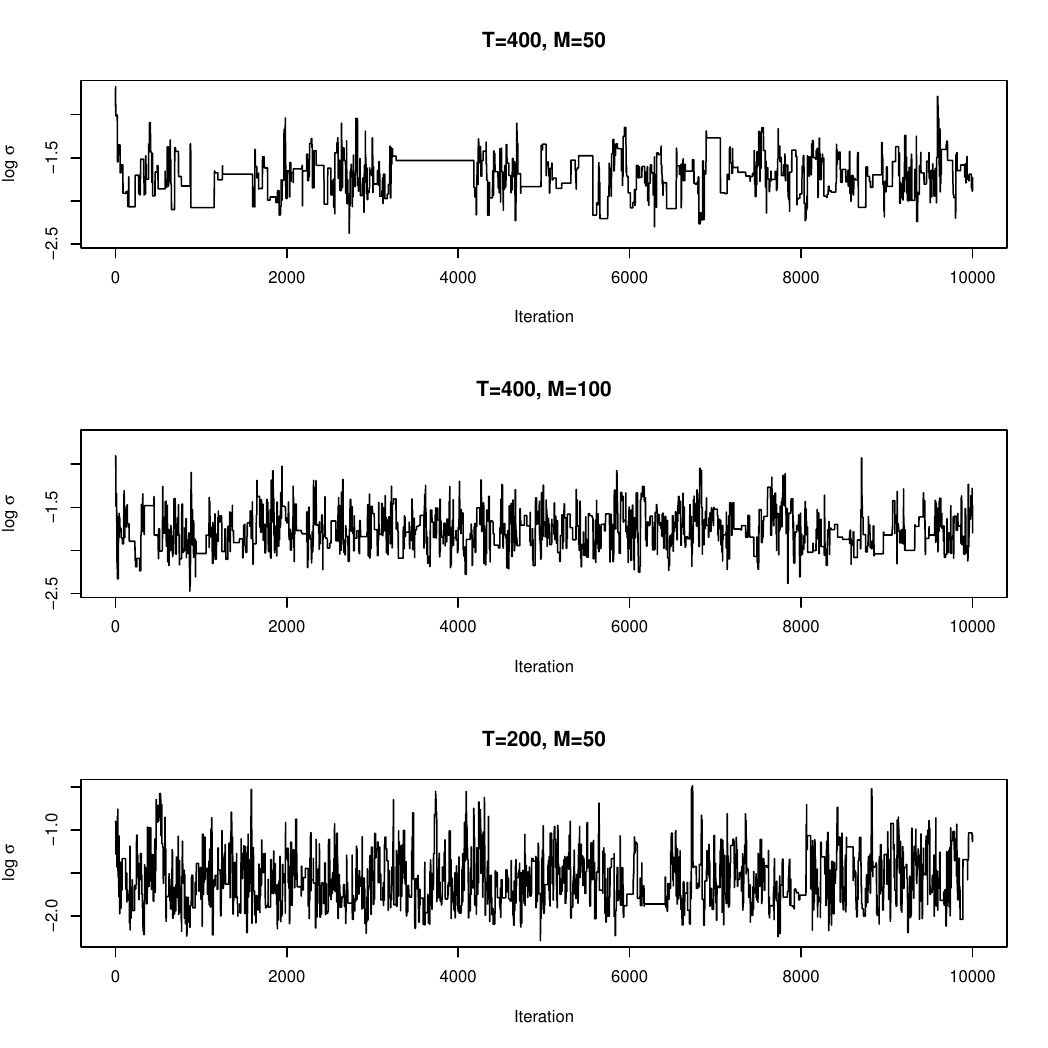}} 
 \caption{\label{Fig:PMMHthree}Trace plots for $\theta_3=\log \sigma$ from the pseudo-marginal random walk Metropolis algorithm using a likelihood estimate derived from a bootstrap particle filter. Top: $T=400$, $M=50$, centre: $T=400$, $M=100$, bottom: $T=200$, $M=50$.}
\end{center}
\end{figure}

Figure \ref{Fig:PMMHthree} shows trace plots for $\theta_3=\log \sigma$ when $M=50$ and $M=100$. The aforementioned stickiness is plainly visible when $M=50$ but, while still present, is much reduced when $M=100$. Where only the first half of the data set is used for inference and $M=50$, similar good behaviour to that with the full data and $M=100$ is obtained. \cite{BerDelDou2014} shows that, subject to conditions, as $T\to \infty$ with $M\propto T$, $\log \PFCSphat(y_{1:T}|\theta;U)$ is asymptotically Gaussian with a variance proportional to $T/M$, so the number of particles required to maintain good behaviour increases in proportion to $T$. As additional verification we estimated $\mathsf{Var}[\log \PFCSphat]$ by repeated evaluation at the true parameter value, obtaining: 0.8 ($M=200$), 1.8 ($M=100$) and 4.4 ($M=50$) with the full data, and 0.4 ($M=200$), 0.8 ($M=100$) and 2.0 ($M=50$) with half of the data. In both cases we would expect little stickiness when $M$ is chosen so that $\mathsf{Var}[\log \PFCSphat]\approx 1$.

\subsection{Particle Marginal Metropolis-Hastings}
\label{sec.particleMH}

As described in Section \ref{sec.pseudo}, inference on the parameters of a state-space model can be performed by using the pseudo-marginal Metropolis-Hastings algorithm with the estimate of the marginal likelihood obtained from a particle filter. Interest may, however, be in the joint distribution of the parameters and the states given all of the data. Despite the suggestion inherent in its name, the particle marginal Metropolis-Hastings (PMMH) algorithm produces samples from $(\theta,x_{1:T})|y_{1:T}$. To describe the algorithm we first introduce the idea of a backwards path.   

\textbf{Backwards path and its proposal probability}: The $k$th weighted particle at time $T$, $x_T^{k}$ arose from the unweighted particle at time $T-1$, $\PFCSxtil_{T-1}^k$. This, in turn, was generated by sampling one of  $x_{T-1}^{1},\dots x_{T-1}^{M}$; say $x_{T-1}^{b_{T-1}^k}$, which itself arose from $\PFCSxtil_{T-2}^{b_{T-1}^k}$. In this way, the ancestry of each of the particles that is present at time $T$ can be traced back to one of the initial samples. Denote the set of ancestors (from the weighted samples) of the particle with index $k$ at time $T$ by $(x_1^{b_1^k},\dots,x_{T-1}^{b_{T-1}^k})$; for notational simplicity we also define $b_T^k=k$. We will call $x_{1:T}^{b^k_{1:T}}:=(x_1^{b_1^k},\dots,x_{T}^{b_{T}^k})$ the path back from $k$. We will consider sampling the path back from $k$ with a probability proportional to $w_T^k$:
\begin{equation}
\label{eqn.kbackwardsprob}
q(k|u,\theta)\equiv q(x_{1:T}^{b_{1:T}^k}|u,\theta)
=
\frac{w_T^k}{\sum_{i=1}^M w_T^i}.
\end{equation}
Given a current parameter value $\theta_i$ and, all of the auxiliary variables, $u_i$, used to run a particle filter at $\theta_i$ (including the 
estimate of the posterior $\PFCSpihat(\theta_i;u_i)$) and a backwards path indexed by $k$, $x_{i,1:T}=x_{1:T}^{b_{1:T}^k}$, the PMMH algorithm proceeds as follows:
\begin{enumerate}
\item Propose a new parameter value $\theta'$ from some $q(\theta'|\theta_i)$.
\item[2a.] Sample $u'$ from $q_*(u'|\theta')$ to obtain  $\PFCSpihat(\theta'|u')$ via \eqref{eqn.ubLikeToPost}.
\item[2b.] Sample a backwards path, indexed by $k'$, from $q(k'|u',\theta')$: $x'_{1:T}=x_{1:T}^{'b_{1:T}^{'k^{'}}}$
\setcounter{enumi}{2}
\item Calculate the Metropolis-Hastings acceptance probability:
\[\alpha_{pm}(\theta_i,u_i;\theta';u')
=
1\wedge \frac{\PFCSpihat(\theta';u')q(\theta_i|\theta')}{\PFCSpihat(\theta_i;u_i)q(\theta'|\theta_i)}.
\]
\item With a probability of $\alpha(\theta_i,u_i;\theta';u')$ set $\theta_{i+1}\leftarrow \theta'$ and $u_{i+1}\leftarrow u'$ and $k_{i+1}\gets k'$ so that $\PFCSpihat(\theta_{i+1};u_{i+1})\gets \PFCSpihat(\theta';u')$ and $x_{i+1,1:T}\gets x_{1:T}'$; otherwise $\theta_{i+1}\gets \theta_i$ and $u_{i+1}\gets u_i$ so that $\PFCSpihat(\theta_{i+1};u_{i+1})=\PFCSpihat(\theta_i;u_i)$ and $x_{i+1,1:T}\gets x_{i,1:T}$.
\end{enumerate}
The only differences from the pseudo-marginal Metropolis-Hastings using the estimator from a particle filter are Step 2b which samples the new path and the part of Step 4 which updates the path. Indeed, the pseudo-marginal algorithm can be viewed as a PMMH algorithm that discards the path samples. 

\textbf{Validity of PMMH}: We now outline why the PMMH algorithm  targets the posterior distribution for $(\theta,x_{1:T})$.  The first step is to itemize the auxiliary variables, $u$, by naming all of the particles' ancestors. 

 The particle $x_t^m$ was sampled from $q(x_t^m|\PFCSxtil_{t-1}^{m})=q(x_t^m|x_{t-1}^k)$ for a specific $k=a_{t-1}^m$, the index of the time $t-1$ ancestor of the particle with index $m$ at time $t$. Let $x_{1:T}^{1:M}$ denote the set of all weighted particles at all times, and let $a_{1:T-1}^{1:M}$ denote the set of ancestor indices. The complete set of auxiliary variables sampled by the particle filter is $u=(x_{1:T}^{1:M},a_{1:T-1}^{1:M})$ and its proposal probability is
\begin{equation}
\label{eqn.define.qstarugtheta}
q_*(u|\theta)
=
\left\{\prod_{m=1}^M q(x_1^m)\right\}
\prod_{t=2}^T \prod_{m=1}^M 
\frac{w_{t-1}^{a_{t-1}^m}}{\sum_{i=1}^Mw_{t-1}^i}
q(x_t^m|x_{t-1}^{a_{t-1}^m}).
\end{equation}
The PMMH algorithm targets the joint posterior of
\[
\widetilde{\pi}(\theta,u,k)
\propto
\pi_0(\theta)
\PFCSphat(y_{1:T}|\theta;u)
q_*(u|\theta) q(k|u,\theta),
\]
because the probability density/mass of being at $(\theta,u,k)$ and proposing and accepting $(\theta',u',k')$ is
\begin{align*}
\widetilde{\pi}(\theta,u,k)
q(\theta'|\theta)q_*(u'|\theta')q(k'|u',\theta')
&\alpha_{pm}(\theta,u;\theta',u')
=\\
&\widetilde{\pi}(\theta',u',k')
q(\theta|\theta')q_*(u|\theta)q(k|u,\theta)
\alpha_{pm}(\theta',u';\theta,u).
\end{align*}
Finally, we show that the marginal distribution for $\theta,k|y_{1:T}$ is proportional to $\pi_0(\theta)p(y_{1:T},x_{1:T}|\theta)$.
Denote by $u_-$ the set of auxiliary variables other than the path back from $k$ (the latter being each $x_t^{b_t^k}$ value and, when $t>1$, its ancestor index). Then, conditional on the path back from $k$, the probability density of $u_-$ is:
\begin{equation}
\label{eqn.qstaruminus}
q_*(u_-|b_{1:T}^k,x_{1:T}^{b_{1:T}^k},\theta)
=
\left\{\prod_{m\ne b^k_1} q(x_1^m)\right\}
\prod_{t=2}^T \prod_{m\ne b^k_t} 
\frac{w_{t-1}^{a_{t-1}^m}}{\sum_{i=1}^Mw_{t-1}^i}
q(x_t^m|x_{t-1}^{a_{t-1}^m}).
\end{equation}
Given \eqref{eqn.qstaruminus}, the form for $q_*(u|\theta)$ in \eqref{eqn.define.qstarugtheta} and that $a_{t-1}^{b_t^k}=b_{t-1}^k$, 
\begin{equation}
\label{eqn.decompose.q}
q_*(u|\theta)= q_*(u_-|b_{1:T}^k,x_{1:T}^{b_{1:T}^k})
\times
q(x_1^{b_1^k})
\prod_{t=2}^T\frac{w_{t-1}^{b_{t-1}^k}}{\sum_{i=1}^M w_{t-1}^i}q(x_t^{b_t^k}|x_{t-1}^{b_{t-1}^k}).
\end{equation}
Then using $\PFCSphat(y_{1:T}|\theta,u)$ from \eqref{eqn.PFphat}, $q(k|u,\theta)$ from \eqref{eqn.kbackwardsprob} and, subsequently, the definitions \eqref{eqn.wOne} and \eqref{eqn.tWeight} of the weights, 
\begin{align}
\nonumber
\widetilde{\pi}(\theta,u,k)
&\propto
\pi_0(\theta)\PFCSphat(y_{1:T}|\theta;u)
q_*(u|\theta) q(k|u,\theta)\\
&= M^{-T}\pi_0(\theta)
\nonumber
q(x_1^{b_1^k})w_1^{b_1^k}\prod_{t=2}^Tq(x_t^{b_t^k}|x_{t-1}^{b_{t-1}^k}) w_t^{b_{t}^k}
\times q_*(u_-|b_{1:T}^k,x_{1:T}^{b_{1:T}^k},\theta)\\
\nonumber
&= M^{-T}\pi_0(\theta)
p(x_1^{b_1^k}|\theta)p(y_1|x_1^{b_1^k},\theta)\prod_{t=2}^{T} p(x_t^{b_t^k}|x_{t-1}^{b_{t-1}^k},\theta)p(y_t|x_t^{b_t^k},\theta)
q_*(u_-|b_{1:T}^k,x_{1:T}^{b_{1:T}^k},\theta)\\
&= M^{-T}\pi_0(\theta)
p(x_{1:T}^{b_{1:T}^k},y_{1:T}|\theta)
q_*(u_-|b_{1:T}^k,x_{1:T}^{b_{1:T}^k},\theta),
\label{eqn.altExtend}
\end{align}
because of the conditional independence structure. 
Integrating out $u_-$ and recalling that the algorithm sets $x_{1:T}=X_{1:T}^{b_{1:T}^k}$, gives the required result.

Additionally, the derivation of \eqref{eqn.altExtend} has shown that
\begin{equation}
\label{eqn.prove.unbiased}
\PFCSphat(y_{1:T}|\theta;u)
q_*(u|\theta) q(k|u,\theta)
=
M^{-T}p(x_{1:T}^{b_{1:T}^k},y_{1:T}|\theta)
q_*(u_-|b_{1:T}^k,x_{1:T}^{b_{1:T}^k},\theta).
\end{equation}
We need to integrate out $u$ and $k$, or, equivalently, $u_-$, $b_{1:T}^k$ and $x_{1:T}^{b^k_{1:T}}$.
After integrating out $u_-$, the indices $b^k_{1:T}$ become irrelevant (so $x_{1:T}^{b_{1:T}^k}$ can be relabelled as  $x_{1:T}$) but we still need to sum over the $M^T$ possibilities.  Thus $q_*(u|\theta) \PFCSphat(y_{1:T}|\theta;u)$ integrates to $p(y_{1:T}|\theta)$; \emph{i.e.}, $\PFCSphat(y_{1:T}|\theta;U)$ is unbiased.


\subsection{Particle Gibbs}
Let us fix $\theta$ for now, and imagine we have a single path: $x_{1:T}^1:= (x_1^{1},\dots,x_T^{1})$. We can run a particle filter \emph{conditional} on the existence of this path. Firstly,  
\begin{enumerate}
    \item[\textbf{Step 0}:] Sample $x_1^{2:M}$ from $q(x_1)$ and weight these particles and $x_1^1$ according to \eqref{eqn.wOne}.
\end{enumerate}
Given a set of weighted particles at time $t-1$, we then iterate the following to obtain a set of weighted particles at time $t$:
\begin{itemize}
\setlength\itemsep{0ex}
    \item[\textbf{Resample}:] Set $\PFCSxtil_{t-1}^1=x_{t-1}^1$. Sample $M-1$ times from the weighted approximation to $p(x_{t-1}|y_{1:t-1},\theta)$, $\left(x_{t-1}^1,w_{t-1}^1\right),\dots,\left(x_{t-1}^M,w_{t-1}^M\right)$, to obtain $\PFCSxtil_{t-1}^{2:M}$.
    \item[\textbf{Propagate}:] We have $x_t^1$. For each $m=2,\dots,M$, sample $x_t^{m}$ from a proposal $q(x_t|\PFCSxtil_{t-1}^{m})$.
    \item[\textbf{Weight}:] For each $m=1,\dots,M$, set 
    $
w_t^{m}
=
{p(y_t|x_t^{m},\theta)p(x_{t}^{m}|\PFCSxtil_{t-1}^{m},\theta)}/
{q(x_t^{m}|\PFCSxtil_{t-1}^{m})}
    $.
\end{itemize}
Finally, we choose a new path (which could be the existing path):
\begin{itemize}
    \item[\textbf{New path}:] Sample a new path $x_{1:T}^{b_{1:T}^{k'}}$ with probability proportional to $w_T^{k'}$ \eqref{eqn.kbackwardsprob}. Relabel $x_t^1\gets x_t^{b_t^{k'}}$, $t=1,\dots,T$. 
\end{itemize}

\begin{figure}
\begin{center}
 \rotatebox{90}
{\includegraphics[height=5in,width=5in,angle=270]{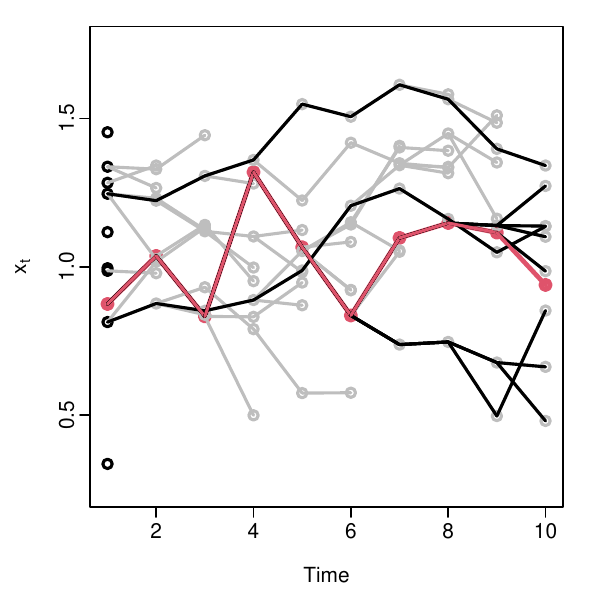}} 
 \caption{\label{Fig:PGibbs1} Schematic of the conditional particle filter step of the Particle Gibbs sampler. The current state is shown in red. We simulate the output of a particle filter conditional on one of the final paths of the filter being the current state. This is simply achieved by setting the first particle at each time-step to be equal to the value of the current state, and for time $t=2,\ldots,T$ descended from the current state at time $t-1$. The other resampling and propagation steps are implemented as per a standard particle filter. The paths that we can resample at time $T$ are shown in black, with particles simulated at earlier times and discarded shown in grey. At time $T$ we resample one of the paths, and that is the new state of our MCMC algorithm. One of the challenges with this step is the particle degeneracy of the particle filter, which can mean that most, if not all, of the paths we can resample will be the same as the current state for small values of $t$.}
\end{center}
\end{figure}

The repeated resample, propagate and weight steps sample from $q_*(u_-|b^k_{1:T},x_{1:T}^{b^k_{1:T}}, \theta)$ in \eqref{eqn.qstaruminus}
with $b_{1:T}^k=(1,\dots,1)$ so $x_{1:T}^{b^k_{1:T}}=x_{1:T}^1$. If $x_{1:T}^1$ is a draw from
$p(x_{1:T}|y_{1:T},\theta)$ then, by \eqref{eqn.altExtend}, the joint distribution of all of the auxiliary variables is
\[
\widetilde{\pi}(u,k)\propto \PFCSphat(y_{1:T}|\theta;u)q_*(u|\theta)q(k=1|u,\theta).
\]
The final step drops the index $k=1$ and the associated previous path and samples a new path by sampling a new $k'$, leading $\widetilde{\pi}(u,k')\propto \PFCSphat(y_{1:T}|\theta;u)q_*(u|\theta)q(k'|u,\theta)$. Using 
\eqref{eqn.altExtend} again this means that the marginal for the new path is proportional to $p(x_{1:T}^{b_{1:T}^{k'}},y_{1:T}|\theta)$, as required.

The algorithm is called \emph{particle Gibbs} as we always accept the sampled path for the state -- although in many cases it will be similar to the current path. This is particularly true for the state at small values of $t$, as depicted in Figure \ref{Fig:PGibbs1} which shows that the number of distinct values of the state we can sample decreases monotonically as we go back in time. We return to this issue in the next section.

Given the new path, $\theta|x_{1:T},y_{1:T}$ can be sampled via any valid Metropolis-Hastings move. In some cases, the conditional posterior for $\theta$, or some subset of its components, is tractable and one or more Gibbs moves may be used. Thus, the particle Gibbs algorithm gives a way of approximately implementing a Gibbs sampler, via an approximation of the update for the hidden state given the parameters. As such, it inherits any dependence between $\theta$ and $x_{1:T}$ that is present in the ideal Gibbs sampler. If there is high dependence then it will mix poorly, like the ideal Gibbs sampler -- and the reparameterisation ideas we introduced earlier should be considered to improve mixing. Alternatively, in these scenarios, the PMMH algorithm may mix better if suitable moves in $\theta$ space are used.

{\bf Example 1: Particle Gibbs}

\begin{figure}
\begin{center}
 \rotatebox{90}
{\includegraphics[height=5in,width=5in,angle=270]{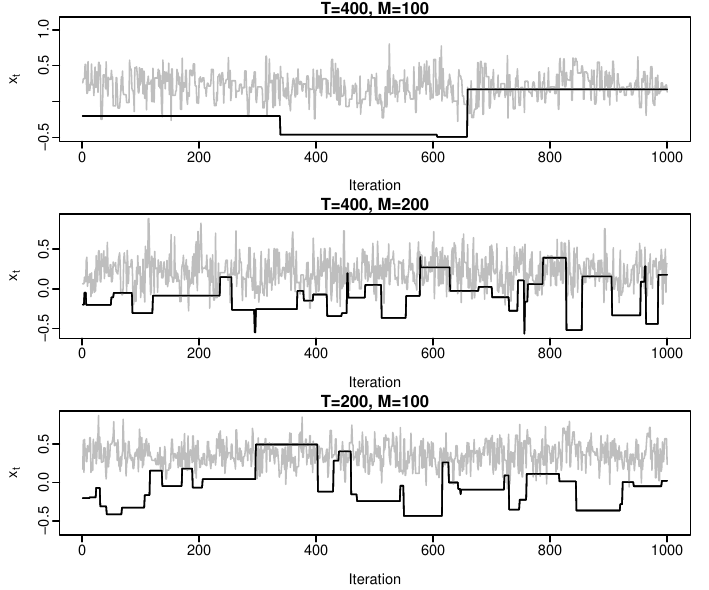}} 
 \caption{\label{Fig:PGibbs2} Trace plots for $x_t$ for the particle Gibbs step for $t=1$ (black) and $t=0.9T$ (grey) for different values of numbers of particles, $M$ and time steps $T$: $(M,T)=(100, 400)$ (top) (200, 400) (middle) and (100, 200) (bottom). }
\end{center}
\end{figure}

To show some of the issues with the particle Gibbs sampler, we investigate the mixing of the particle Gibbs step for updating the hidden state in the stochastic volatility model using the same data set as in Section \ref{sec.pseudo}. We do this for $\theta$ fixed at the truth; that is, we focus solely on how well the particle Gibbs step performs, and do not consider the mixing of $\theta$ given the hidden state, that is common to other Gibbs or Metropolis-within-Gibbs algorithms. 

Figure \ref{Fig:PGibbs2} shows trace plots of $x_t$ for two different values of $t$, repeated for different values of the number of particles $M$ and length of time $T$. We highlight three points. First, mixing is worse for smaller values of $t$, and this is linked to the degeneracy of the paths in the particle filter that we discussed above. Second, care is needed to choose $M$ appropriately to get the particle Gibbs update of the state to mix well -- too small a value will lead to poorly mixing states for small $t$. Finally, as with PMMH, we see evidence that we need to increase $M$ as $T$ increases, and again a rule-of-thumb is that we should increase $M$ linearly with $T$.

\subsection{Extensions and Improvements for Particle MCMC}


There have been a number of approaches to improving the efficiency of particle MCMC; we highlight three. First, for the PMMH we saw that the chain can be sticky if we have a high-variance estimator of the posterior density. One approach to avoid this is to increase the number of particles. However, \cite{deligiannidis2018correlated} introduced the use of correlated estimators of the posterior density within the pseudo-marginal algorithm. The idea is that when we propose a new value, $\theta'$, the driving noise we use to sample the auxiliary variables $U'$ is a perturbation of (rather than independent of) that used to sample $U$. Thus, our estimator $\widehat{\pi}(\theta';U')$ of $\pi(\theta')$ is, typically, positively correlated with the estimator $\widehat{\pi}(\theta;U)$ of $\pi(\theta)$,  so that the ratio $\widehat{\pi}(\theta';U')/\widehat{\pi}(\theta;U)$ that appears in the accept-reject probability has a lower variance. This improves mixing, though constructing correlated estimators of $\pi(\theta)$ for different $\theta$ values for the particle filter is challenging due to the discrete nature of the resampling step \cite[see e.g.][]{sen2018coupling}.

As explained earlier, for the Particle Gibbs algorithm, when $T$ is large compared with $M$, the early ancestors of every $x_T^m$, $m=1,\dots,M$, are usually identical, so that they rarely change from one iteration to the next. 

The backwards sampling of \cite{whiteley2010discussion} and the ancestor sampling of \cite{LinJorSch2014} offer  
mitigation against path degeneracy. The two methods are probabilistically equivalent when applied to hidden Markov models, and, in essence, at each time point $t-1$, they allow the reference path to date, $x^1_{1:t-1}$,  to, perhaps, be changed to the path to date of one of the other particles by sampling from the set of $M$ such paths with a probability that takes into account how well each of the $M$ paths to date fits with the rest of the reference path. Specifically, we replace the  Propagate step of the Particle Gibbs algorithm with:

\textbf{Propagate}: We have $x^1_t$. For each $m=2,\dots,M$, sample $x_t^{m}$ from a proposal $q(x_t|\PFCSxtil_{t-1}^{m})$.  \emph{Also} choose a new $x^1_{1:t-1}$ by sampling $J$ from $\{1,\dots,M\}$ with
\[
\mathbb{P}(J=j)
=\frac{w_{t-1}^j p(x^1_t|x^j_{t-1},\theta)}{\sum_{m=1}^Mw_{t-1}^m p(x^1_t|x^m_{t-1},\theta)},
\]
and setting $x^1_{1:t-1}$ to be the 
 backward path from (and including) $x^J_{t-1}$.
The weight $w_{t-1}^j$ can be thought of as the prior for path $j$, and $p(x^1_t|x^j_{t-1},\theta)$ is the likelihood for the next point in the path, $x_t^1$, given the path to date, $x_{1:t-1}^j$.  

This is most natural to implement in cases where $p(x_t|x_{t-1},\theta)$ is tractable; however, it is possible to extend it to more general hidden processes such as diffusions and Markov jump processes; see \cite{LinJorSch2014}.

\begin{figure}
\begin{center}
 \rotatebox{90}
{\includegraphics[height=4in,width=6in,angle=270]{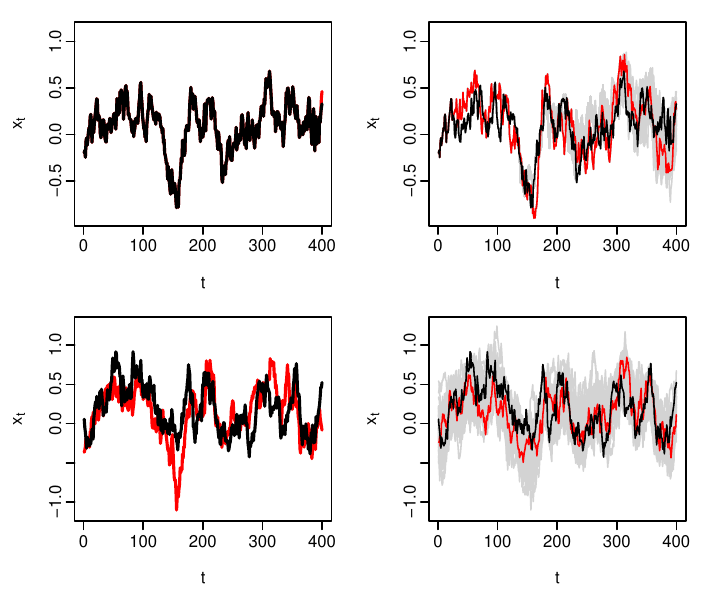}} 
 \caption{\label{Fig:PGibbs_anc} Plots of the path of the state, $x_t$, for particle Gibbs, applied to the stochastic volatility example, both with ancestor sampling (bottom row) and without  (top row). For both cases, the left-hand plots show the initial path (black) and the path sampled after one iteration (red); whilst the right-hand plots show the initial path (black), paths every iteration (grey), and the final path after fifty iterations (red). Results are for $T=400$ and $M=100$.}
\end{center}
\end{figure}

To see the benefit of ancestor sampling, we return to the stochastic volatility example and compare the particle Gibbs step for updating the path of the state both with and without ancestor sampling. See Figure \ref{Fig:PGibbs_anc} and compare the mixing as shown in the top row, which is for standard particle Gibbs, and for the bottom row, which is for particle Gibbs with ancestor sampling. Whilst they have similar mixing of the state for $t$ close to the terminal time, $T$; ancestor sampling avoids the path degeneracy that means standard particle Gibbs mixes very slowly for small $t$. The advantages of ancestor sampling has been shown theoretically, with \cite{karjalainen2023mixing} showing that convergence of particle Gibbs with ancestor sampling can occur in $O(\log T)$ iterations with a number of particles $M$ that does not need to grow with $T$. 

More recently, \cite{Malory2021} (see  \cite{FinThi2023} for a related alternative) samples $u_-$ using perturbations of the current path $x^1_{1:T}$ in such a way that the set of all backward paths arises from an exchangeable distribution. For example, if simulating $X_t|X_{t-1},\theta$ involves simulating $Z_{t-1}\sim \mathsf{N}(0,1)$ then, given the value for the reference path, $Z_{t-1}^1$, the values for all other paths are simulated jointly from the  conditional implied by the exchangeable distribution\[
\begin{bmatrix}
    Z^1_{t-1}\\
    Z^2_{t-1}\\
    \dots\\
    Z^{M}_{t-1}
\end{bmatrix}
\sim
\mathsf{N}_M\left(\underline{0},
\begin{bmatrix}
    1&\rho&\dots&\rho\\
    \rho&1&\dots&\rho\\
    \dots&\dots&\dots&\dots\\
    \rho&\rho&\dots&1
\end{bmatrix}
\right).
\]
When $\rho$ is close to $1$, $X_t^{2:M}$ are, typically, close to $X_t^1$ and, more generally, the new paths  are more similar to the reference path than with standard Particle Gibbs, which simulates $Z_{t-1}^{2:M}\stackrel{iid}{\sim}\mathsf{N}(0,1)$; this leads to multiple distinct backwards paths. \cite{FinThi2023} enhances the performance still further by also incorporating  ancestor sampling. 

The other challenge with the particle Gibbs algorithm is that it can mix slowly if the underlying Gibbs sampler mixes poorly due to strong dependence between the parameter and the hidden state. However, the particle Gibbs sampler gives flexibility as to what part of $(\theta,x_{1:T})$ is conditioned on and what is updated within the particle Gibbs update. Thus one option is to partially update $\theta$ within the particle Gibbs update. Ideas for doing so are described in \cite{fearnhead2016augmentation}.


\section{Discussion}

This chapter has provided an introduction to MCMC methods for state-space models. There are three main aspects that have been covered. Firstly, if there is strong, or long-range, dependence in the state-space model, then an efficient MCMC algorithm will need to update blocks of the state-process in a single move. We have looked at examples where efficient block updates are straightforward to construct; however, in many applications this can be difficult. Secondly, strong correlation between the parameters and the state process can lead to slow mixing of the MCMC algorithm, even if there are efficient methods for updating the state process. To improve mixing, either reparameterisation of the model, or joint updates of the state and the parameters are needed. Finally, particle MCMC supplies a general class of methods for constructing efficient MCMC moves for state-space models. Typically, the computational cost of these increases quadratically with the number of observation times, which is much better than the exponential-in-time degradation that is often observed when using importance samplers for the hidden-state.  Furthermore, particle Gibbs with ancestor sampling can have a better scaling of the computational cost with the number of observation times \cite[]{karjalainen2023mixing}.

While we have focussed on MCMC methods and particle MCMC that is specifically designed for Bayesian inference for state-space models, it is possible to use generic MCMC algorithms to sample from the joint posterior of the parameters and the states. As this joint posterior is often high-dimensional, gradient-based MCMC, such as the Metropolis-adjusted Langevin algorithm \cite[]{xifara2014langevin} or Hamiltonian Monte Carlo (see Chapter 5), are to be preferred. Such algorithms can be implemented using general MCMC software such as STAN \cite[]{carpenter2017stan} or PyMC \cite[]{patil2010pymc}, or through more refined implementations \cite[e.g.][]{kleppe2019dynamically}. However the methods we have described, particularly the particle MCMC algorithms, can deal with discrete state spaces or models where we can simulate from the state process but not analytically calculate the state's transition densities; these are models where generic gradient-based MCMC would not work.


While we have focused on discrete-time state-processes, many of the issues extend naturally to continuous-time processes. For example, the issue of model parameterisation for diffusion models is discussed in \citet{Roberts/Stramer:2001}, and for these models certain parameterisations can lead to MCMC algorithms which are reducible. Extensions of the Forward-Backward algorithm to continuous-time models is considered in \citet{Fearnhead/Meligkotsidou:2004}, and independence sampler updates for the state-process in diffusion models are developed in \citet{golightly2008bayesian}. Also, whilst we have described Particle filter methods in terms of their use within particle MCMC for sampling from the posterior distribution, the same ideas and methods can be useful for non-Bayesian analyses. For example, to implement Monte Carlo EM algorithms \cite[]{kantas2015particle} or to estimate the gradient of the marginal likelihood \cite[e.g.][]{poyiadjis2011particle,nemeth2016particle}, that is the gradient of the likelihood after integrating out the unobserved latent state process, which can then be used within gradient ascent algorithms to find the maximum likelihood estimator.

\textbf{Acknowledgements}: the authors are particularly grateful to Samuel Holdstock and Lanya Yang for their help in proof reading this chapter.

\bibliographystyle{apalike} 
\bibliography{thesis,coal}

\begin{thebibliography}{}

\bibitem[Andrieu et~al., 2010]{AndDouHol2010}
Andrieu, C., Doucet, A., and Holenstein, R. (2010).
\newblock Particle {M}arkov chain {M}onte {C}arlo methods.
\newblock {\em Journal of the Royal Statistical Society: Series B},
  72(3):269--342.

\bibitem[Andrieu and Roberts, 2009]{AndRob2009}
Andrieu, C. and Roberts, G.~O. (2009).
\newblock {The pseudo-marginal approach for efficient Monte Carlo
  computations}.
\newblock {\em The Annals of Statistics}, 37(2):697 -- 725.

\bibitem[Andrieu and Thoms, 2008]{Andrieu/Thoms:2008}
Andrieu, C. and Thoms, J. (2008).
\newblock A tutorial on adaptive {MCMC}.
\newblock {\em Statistics and Computing}, 18(4):343--373.

\bibitem[Andrieu and Vihola, 2015]{AndVih2015}
Andrieu, C. and Vihola, M. (2015).
\newblock Convergence properties of pseudo-marginal {M}arkov chain {M}onte
  {C}arlo algorithms.
\newblock {\em The Annals of Applied Probability}, 25(2):1030--1077.

\bibitem[Ball and Rice, 1992]{Ball:1992}
Ball, F.~G. and Rice, J.~A. (1992).
\newblock Stochastic models for ion channels: Introduction and bibliography.
\newblock {\em Mathematical Biosciences}, 112:189--206.

\bibitem[Boys et~al., 2000]{Boys/Henderson/Wilkinson:2000}
Boys, R.~J., Henderson, D.~A., and Wilkinson, D.~J. (2000).
\newblock Detecting homogeneous segments in {DNA} sequences by using hidden
  {M}arkov models.
\newblock {\em Journal of the Royal Statistical Society, Series C},
  49:269--285.

\bibitem[Bérard et~al., 2014]{BerDelDou2014}
Bérard, J., Moral, P.~D., and Doucet, A. (2014).
\newblock A lognormal central limit theorem for particle approximations of
  normalizing constants.
\newblock {\em Electron. J. Probab.}, 19:no. 93, 1--28.

\bibitem[Carpenter et~al., 2017]{carpenter2017stan}
Carpenter, B., Gelman, A., Hoffman, M.~D., Lee, D., Goodrich, B., Betancourt,
  M., Brubaker, M.~A., Guo, J., Li, P., and Riddell, A. (2017).
\newblock Stan: A probabilistic programming language.
\newblock {\em Journal of Statistical Software}, 76(1).

\bibitem[Carter and Kohn, 1994]{Carter/Kohn:1994}
Carter, C.~K. and Kohn, R. (1994).
\newblock {On Gibbs sampling for state space models}.
\newblock {\em Biometrika}, 81(3):541--553.

\bibitem[Chopin and Papaspiliopoulos, 2020]{chopin2020introduction}
Chopin, N. and Papaspiliopoulos, O. (2020).
\newblock {\em An introduction to sequential {Monte Carlo}}.
\newblock Springer.

\bibitem[Deligiannidis et~al., 2018]{deligiannidis2018correlated}
Deligiannidis, G., Doucet, A., and Pitt, M.~K. (2018).
\newblock The correlated pseudomarginal method.
\newblock {\em Journal of the Royal Statistical Society Series B: Statistical
  Methodology}, 80(5):839--870.

\bibitem[Didelot et~al., 2007]{Didelot:2007}
Didelot, X., Achtman, M., Parkhill, J., Thomson, N.~R., and Falush, D. (2007).
\newblock A bimodal pattern of relatedness between the {\it salmonella}
  {Paratyphi A and Typhi} genomes: {Convergence} or divergence by homologous
  recombination?
\newblock {\em Genome Research}, 17:61--68.

\bibitem[Doucet et~al., 2015]{DouPitDelKoh2015}
Doucet, A., Pitt, M.~K., Deligiannidis, G., and Kohn, R. (2015).
\newblock {Efficient implementation of Markov chain Monte Carlo when using an
  unbiased likelihood estimator}.
\newblock {\em Biometrika}, 102(2):295--313.

\bibitem[Fearnhead, 2006]{Fearnhead:2006SC}
Fearnhead, P. (2006).
\newblock Exact and efficient inference for multiple changepoint problems.
\newblock {\em Statistics and Computing}, 16:203--213.

\bibitem[Fearnhead, 2008]{Fearnhead:2008}
Fearnhead, P. (2008).
\newblock Computational methods for complex stochastic systems: A review of
  some alternatives to {MCMC}.
\newblock {\em Statistics and Computing}, 18:151--171.

\bibitem[Fearnhead and Meligkotsidou, 2004]{Fearnhead/Meligkotsidou:2004}
Fearnhead, P. and Meligkotsidou, L. (2004).
\newblock Exact filtering for partially-observed continuous-time {M}arkov
  models.
\newblock {\em Journal of the Royal Statistical Society, series B},
  66:771--789.

\bibitem[Fearnhead and Meligkotsidou, 2016]{fearnhead2016augmentation}
Fearnhead, P. and Meligkotsidou, L. (2016).
\newblock Augmentation schemes for particle {MCMC}.
\newblock {\em Statistics and Computing}, 26:1293--1306.

\bibitem[Fearnhead and Sherlock, 2006]{Fearnhead/Sherlock:2006}
Fearnhead, P. and Sherlock, C. (2006).
\newblock Bayesian analysis of {Markov} modulated {Poisson} processes.
\newblock {\em Journal of the Royal Statistical Society, Series B},
  68:767--784.

\bibitem[Finke and Thiery, 2023]{FinThi2023}
Finke, A. and Thiery, A.~H. (2023).
\newblock {Conditional sequential Monte Carlo in high dimensions}.
\newblock {\em The Annals of Statistics}, 51(2):437 -- 463.

\bibitem[Gelfand et~al., 1995]{Gelfand/Sahu/Carlin:1995}
Gelfand, A.~E., Sahu, S., and Carlin, B.~P. (1995).
\newblock Efficient parameterisations for normal linear mixed models.
\newblock {\em Biometrika}, 82:479--488.

\bibitem[Golightly and Wilkinson, 2008]{golightly2008bayesian}
Golightly, A. and Wilkinson, D.~J. (2008).
\newblock Bayesian inference for nonlinear multivariate diffusion models
  observed with error.
\newblock {\em Computational Statistics \& Data Analysis}, 52(3):1674--1693.

\bibitem[Gordon et~al., 1993]{GSS1993}
Gordon, N., Salmond, D., and Smith, A. (1993).
\newblock Novel approach to nonlinear/non-{G}aussian {B}ayesian state
  estimation.
\newblock {\em {IEE} Proc. F Radar Signal Process. {UK}}, 140(2):107--113.

\bibitem[Harvey, 1989]{Harvey:1989}
Harvey, A.~C. (1989).
\newblock {\em Forecasting, stuctural time series and the {K}alman filter}.
\newblock Cambridge University Press, Cambridge, UK.

\bibitem[Hobert, 2011]{Hobert:2008}
Hobert, J.~P. (2011).
\newblock The data augmentation algorithm: theory and methodology.
\newblock In {\em Handbook of Markov Chain Monte Carlo}. CRC, London.

\bibitem[Hodgson, 1999]{Hodgson:1999}
Hodgson, M. E.~A. (1999).
\newblock A {B}ayesian restoration of an ion channel signal.
\newblock {\em Journal of the Royal Statistical Society, Series B}, 61:95--114.

\bibitem[Hull and White, 1987]{Hull/White:1987}
Hull, J. and White, A. (1987).
\newblock The pricing of options on assets with stochastic volatilities.
\newblock {\em Journal of Finance}, 42:281--300.

\bibitem[Juang and Rabiner, 1991]{Juang/Rabiner:1991}
Juang, B.~H. and Rabiner, L.~R. (1991).
\newblock Hidden {M}arkov models for speech recognition.
\newblock {\em Technometrics}, 33:251--272.

\bibitem[Jungbacker and Koopman, 2007]{Jungbacker/Koopman:2007}
Jungbacker, B. and Koopman, S.~J. (2007).
\newblock {Monte Carlo Estimation for Nonlinear Non-Gaussian State Space
  Models}.
\newblock {\em Biometrika}, 94:827--839.

\bibitem[Kalman and Bucy, 1961]{Kalman/Bucy:1961}
Kalman, R. and Bucy, R. (1961).
\newblock New results in linear filtering and prediction theory.
\newblock {\em Journal of Basic Engineering, Transacation ASME series D},
  83:95--108.

\bibitem[Kantas et~al., 2015]{kantas2015particle}
Kantas, N., Doucet, A., Singh, S.~S., Maciejowski, J., and Chopin, N. (2015).
\newblock On particle methods for parameter estimation in state-space models.
\newblock {\em Statistical Science}, 30:328--351.

\bibitem[Karjalainen et~al., 2023]{karjalainen2023mixing}
Karjalainen, J., Lee, A., Singh, S.~S., and Vihola, M. (2023).
\newblock Mixing time of the conditional backward sampling particle filter.
\newblock arXiv:2312.17572.

\bibitem[King, 2014]{king2014statistical}
King, R. (2014).
\newblock Statistical ecology.
\newblock {\em Annual Review of Statistics and its Application}, 1:401--426.

\bibitem[Kitagawa, 1987]{Kitagawa:1987}
Kitagawa, G. (1987).
\newblock {Non-Gaussian state-space modelling of non-stationary time series
  (with discussion)}.
\newblock {\em Journal of the American Statistical Association}, 82:1032--1063.

\bibitem[Kleppe, 2019]{kleppe2019dynamically}
Kleppe, T.~S. (2019).
\newblock {Dynamically rescaled Hamiltonian Monte Carlo for Bayesian
  hierarchical models}.
\newblock {\em Journal of Computational and Graphical Statistics},
  28(3):493--507.

\bibitem[Kong et~al., 1994]{kong1994sequential}
Kong, A., Liu, J.~S., and Wong, W.~H. (1994).
\newblock Sequential imputations and {B}ayesian missing data problems.
\newblock {\em Journal of the American Statistical Association},
  89(425):278--288.

\bibitem[Lekone and Finkenstädt, 2006]{LekFin2006}
Lekone, P.~E. and Finkenstädt, B.~F. (2006).
\newblock Statistical inference in a stochastic epidemic {SEIR} model with
  control intervention: Ebola as a case study.
\newblock {\em Biometrics}, 62(4):1170--1177.

\bibitem[Lindsten et~al., 2014]{LinJorSch2014}
Lindsten, F., Jordan, M.~I., and Sch{{\"o}}n, T.~B. (2014).
\newblock Particle {G}ibbs with ancestor sampling.
\newblock {\em Journal of Machine Learning Research}, 15(63):2145--2184.

\bibitem[Liu, 1994]{Liu:1994}
Liu, J.~S. (1994).
\newblock Fraction of missing information and convergence rate of data
  augmentation.
\newblock In {\em Computing Science and Statistics: Proc. 26th Symposium on the
  Interface}, pages 490--496. Interface Foundation of North America, Fairfax
  Station, VA.

\bibitem[Malory, 2021]{Malory2021}
Malory, S. (2021).
\newblock {\em Bayesian inference for stochastic processes}.
\newblock PhD thesis, Department of Mathematics and Statistics.
\newblock Lancaster University.

\bibitem[Nemeth et~al., 2016]{nemeth2016particle}
Nemeth, C., Fearnhead, P., and Mihaylova, L. (2016).
\newblock Particle approximations of the score and observed information matrix
  for parameter estimation in state--space models with linear computational
  cost.
\newblock {\em Journal of Computational and Graphical Statistics},
  25(4):1138--1157.

\bibitem[Papaspiliopoulos, 2003]{Papaspiliopoulos:2003}
Papaspiliopoulos, O. (2003).
\newblock {\em Non-centered parameterizations for hierarchical models and data
  augmentation}.
\newblock PhD thesis, Dept. Mathematics and Statistics, Lancaster University.

\bibitem[Papaspiliopoulos et~al., 2003]{Papaspiliopoulos/Roberts/Skold:2003}
Papaspiliopoulos, O., Roberts, G.~O., and Sk\"{o}ld, M. (2003).
\newblock {Non-centred parameterisations for hierarchical models and data
  augmentation (with discussion)}.
\newblock In Bernardo, J.~M., Bayarri, M.~J., Berger, J.~O., Dawid, A.~P.,
  Heckerman, D., Smith, A. F.~M., and West, M., editors, {\em {Bayesian
  Statistics 7}}, London. Clarendon Press.

\bibitem[Papaspiliopoulos et~al., 2007]{Papaspiliopoulos/Roberts/Skold:2007}
Papaspiliopoulos, O., Roberts, G.~O., and Sk\"{o}ld, M. (2007).
\newblock A general framework for the parameterization of hierarchical models.
\newblock {\em Statistical Science}, 22:59--73.

\bibitem[Patil et~al., 2010]{patil2010pymc}
Patil, A., Huard, D., and Fonnesbeck, C.~J. (2010).
\newblock {PyMC: Bayesian stochastic modelling in Python}.
\newblock {\em Journal of Statistical Software}, 35(4):1--81.

\bibitem[Pitt et~al., 2012]{PitSilGioKoh2012}
Pitt, M.~K., dos Santos~Silva, R., Giordani, P., and Kohn, R. (2012).
\newblock {On some properties of Markov chain Monte Carlo simulation methods
  based on the particle filter}.
\newblock {\em Journal of Econometrics}, 171(2):134--151.

\bibitem[Pitt and Shephard, 1999]{Pitt/Shephard:1999}
Pitt, M.~K. and Shephard, N. (1999).
\newblock {Filtering via simulation: auxiliary particle filters}.
\newblock {\em Journal of the American Statistical Association}, 94:590--599.

\bibitem[Poyiadjis et~al., 2011]{poyiadjis2011particle}
Poyiadjis, G., Doucet, A., and Singh, S.~S. (2011).
\newblock Particle approximations of the score and observed information matrix
  in state space models with application to parameter estimation.
\newblock {\em Biometrika}, 98(1):65--80.

\bibitem[Rabiner and Juang, 1986]{Rabiner:1986}
Rabiner, L.~R. and Juang, B.~H. (1986).
\newblock An introduction to hidden {M}arkov models.
\newblock {\em IEEE ASSP Magazine}, pages 4--15.

\bibitem[Roberts et~al., 2004]{Roberts/Papaspiliopoulos/Dellaportas:2004}
Roberts, G.~O., Papaspiliopoulos, O., and Dellaportas, P. (2004).
\newblock Bayesian inference for non-{G}aussian {O}rnstein-{U}hlenbeck
  stochastic volatility processes.
\newblock {\em Journal of the Royal Statistical Society, series B},
  66:369--393.

\bibitem[Roberts and Rosenthal, 2009]{Roberts/Rosenthal:2006}
Roberts, G.~O. and Rosenthal, J.~S. (2009).
\newblock Examples of adaptive {MCMC}.
\newblock {\em Journal of Computational and Graphical Statistics},
  18(2):349--367.

\bibitem[Roberts and Sahu, 1997]{Roberts/Sahu:1997}
Roberts, G.~O. and Sahu, S.~K. (1997).
\newblock Updating schemes, correlation structure, blocking and
  parameterization for the {G}ibbs sampler.
\newblock {\em Journal of the Royal Statistical Society, Series B},
  59:291--317.

\bibitem[Roberts and Stramer, 2001]{Roberts/Stramer:2001}
Roberts, G.~O. and Stramer, O. (2001).
\newblock On inference for partially observed nonlinear diffusion models using
  the {Metropolis-Hastings} algorithm.
\newblock {\em Biometrika}, 88:603--621.

\bibitem[Rue and Held, 2005]{Rue/Held:2005}
Rue, H. and Held, L. (2005).
\newblock {\em {Gaussian Markov Random Fields: Theory and Applications}}.
\newblock CRC Press/Chapman and Hall.

\bibitem[Scott, 2002]{Scott:2002a}
Scott, S.~L. (2002).
\newblock Bayesian methods for hidden {M}arkov models: {R}ecursive computing in
  the 21st century.
\newblock {\em Journal of the American Statistical Association}, 97:337--351.

\bibitem[Sen et~al., 2018]{sen2018coupling}
Sen, D., Thiery, A.~H., and Jasra, A. (2018).
\newblock On coupling particle filter trajectories.
\newblock {\em Statistics and Computing}, 28:461--475.

\bibitem[Shephard, 1996]{Shephard:1996}
Shephard, N. (1996).
\newblock {Statistical aspects of ARCH and stochastic volatility}.
\newblock In Cox, D.~R., Hinkley, D.~V., and Barndorff-Nielsen, O.~E., editors,
  {\em Time Series Models in Econometrics, Finance and Other Fields}, pages
  1--67, Chapman and Hall, London.

\bibitem[Shephard and Pitt, 1997]{Shephard/Pitt:1997}
Shephard, N. and Pitt, M.~K. (1997).
\newblock Likelihood analysis of non-{G}aussian measurement time series.
\newblock {\em Biometrika}, 84:653--667.

\bibitem[Sherlock et~al., 2010]{Sherlock/Fearnhead:2008}
Sherlock, C., Fearnhead, P., and Roberts, G.~O. (2010).
\newblock The random walk {M}etropolis: Linking theory and practice through a
  case study.
\newblock {\em Statistical Science}, 25(2):172--190.

\bibitem[Sherlock et~al., 2017]{SheThiLee2017}
Sherlock, C., Thiery, A.~H., and Lee, A. (2017).
\newblock {Pseudo-marginal Metropolis–Hastings sampling using averages of
  unbiased estimators}.
\newblock {\em Biometrika}, 104(3):727--734.

\bibitem[Sherlock et~al., 2015]{SheThiRobRos2015}
Sherlock, C., Thiery, A.~H., Roberts, G.~O., and Rosenthal, J.~S. (2015).
\newblock {On the efficiency of pseudo-marginal random walk Metropolis
  algorithms}.
\newblock {\em The Annals of Statistics}, 43(1):238 -- 275.

\bibitem[Smith and Santos, 2006]{Smith/Santos:2006}
Smith, J.~Q. and Santos, A.~F. (2006).
\newblock Second order filter distribution approximations for financial time
  series with extreme outlier.
\newblock {\em Journal of Business and Economic Statistics}, 24:329--337.

\bibitem[Touloupou et~al., 2020]{touloupou2020scalable}
Touloupou, P., Finkenst{\"a}dt, B., and Spencer, S.~E. (2020).
\newblock {Scalable Bayesian inference for coupled hidden Markov and
  semi-Markov models}.
\newblock {\em Journal of Computational and Graphical Statistics},
  29(2):238--249.

\bibitem[Whiteley, 2010]{whiteley2010discussion}
Whiteley, N. (2010).
\newblock Discussion on the paper by {A}ndrieu, {D}oucet and {H}olenstein.
\newblock {\em Journal of the Royal Statistical Society, Series B},
  72(3):306--307.

\bibitem[Wilkinson, 2018]{wilkinson2018stochastic}
Wilkinson, D.~J. (2018).
\newblock {\em Stochastic Modelling for Systems Biology}.
\newblock CRC press.

\bibitem[Xifara et~al., 2014]{xifara2014langevin}
Xifara, T., Sherlock, C., Livingstone, S., Byrne, S., and Girolami, M. (2014).
\newblock Langevin diffusions and the {Metropolis-adjusted Langevin algorithm}.
\newblock {\em Statistics \& Probability Letters}, 91:14--19.

\end{thebibliography}
\end{doublespace}
\end{document}